\documentclass{article}[11]
\usepackage{graphicx}
\usepackage{subfigure}
\usepackage{amsfonts}
\usepackage{amsmath}
\usepackage{amssymb}
\usepackage{makecell}
\usepackage{url}
\usepackage[misc]{ifsym}
\usepackage{fancyhdr}
\usepackage{indentfirst}
\usepackage{enumerate}
\usepackage{cancel}
\usepackage{caption}
\usepackage{dsfont}
\usepackage[colorlinks=true,citecolor=blue, hypertexnames=false]{hyperref}
\usepackage{amsthm}
\usepackage{authblk}
\usepackage{bm}
\usepackage{color}
\usepackage{natbib}
\usepackage{comment}
\usepackage{capt-of}
\usepackage{multirow}
\usepackage{booktabs}
\def\laweq{\buildrel \mathrm{d} \over =}
\newcommand{\E}{\mathbb{E}}
\newcommand{\p}{\mathbb{P}}
\newcommand{\R}{\mathbb{R}}
\newcommand{\N}{\mathbb{N}}

\newcommand{\st}{\mathrm{st}}

\newcommand{\BP}{\mathrm{Beta\ Prime}}
\newcommand{\Ga}{\mathrm{Gamma}}

\renewcommand{\ge}{\geqslant}
\renewcommand{\le}{\leqslant}
\renewcommand{\geq}{\geqslant}
\renewcommand{\leq}{\leqslant}
\renewcommand{\epsilon}{\bm{a}repsilon}

\theoremstyle{plain}
\newtheorem{theorem}{Theorem}
\newtheorem{corollary}{Corollary}
\newtheorem{lemma}{Lemma}
\newtheorem{proposition}{Proposition}
\theoremstyle{definition}
\newtheorem{definition}{Definition}
\newtheorem{example}{Example}
\theoremstyle{remark}
\newtheorem{remark}{Remark}

\newcommand{\esssup}{\mathrm{ess\mbox{-}sup}}
\newcommand{\essinf}{\mathrm{ess\mbox{-}inf}}
\newcommand{\VaR}{\mathrm{VaR}}

\newcommand{\id}{\mathds{1}}

\usepackage{setspace}

\begin{document}

\title{Risk sharing for background risk models with infinite mean}
\author[1]{Ziting Miao}
\author[1]{Yuyu Chen}
\author[1]{Han Li}
\affil[1]{Centre for Actuarial Studies, Department of Economics,
          The University of Melbourne, Australia}
\renewcommand{\Affilfont}{\small}
\date{\today}
\maketitle

\begin{abstract}
Risk sharing is usually viewed as beneficial for risk-averse agents facing independent risks with finite means, but this situation can be reversed for infinite-mean losses. In this paper, we explore the effects of risk sharing for positively dependent infinite-mean losses using background risk models. Under both linear and conditional mean risk sharing, we show that risk sharing of infinite-mean losses can increase the risk in the sense of usual stochastic order. Our results cover classical multivariate loss distributions, including the Pareto-Clayton and the Fr\'echet-Gumbel models. For linear risk sharing, we further derive stochastic dominance relations between diversified portfolios, including those with Pareto marginals coupled by convex mixtures of commonly used Archimedean copulas. For conditional mean risk sharing, we additionally discuss peakedness and stochastic order relations for stable distributions. Consequently, under the stated conditions, retaining the original risk or a less diversified portfolio is preferred under every law-invariant monotone risk measure that is well defined for infinite-mean models.

\end{abstract}

\section{Introduction}
\label{sec:intro}
Many operational decisions involve risks whose extreme values can dominate the aggregate outcomes, known as the  principle of a single big jump in the Extreme Value Theory; see, e.g., \cite{EKM97}. Such phenomena often arise in the realm of infinite-mean models. Examples include natural disaster losses \citep{ibragimov2009nondiversification}, banking operational risks \citep{moscadelli2004modelling}, cyber losses \citep{eling2019actual}, corporate fraud \citep{cheynel2024fraud}, system reliability and failure recovery \citep{asmussen2008asymptotic}, and pandemic severity \citep{cirillo2020tail}. A broader review of infinite-mean models and their applications is provided by \citet{chen2025infinite}.

In this paper, we study how linear risk sharing (LRS) and conditional mean risk sharing (CMRS) affect the risk of infinite-mean losses. Suppose that there are $n$ agents in a risk sharing pool and each agent faces an initial random loss $a_iX_i$ where $a_i>0$ is the exposure.  Under LRS, the shared loss of agent $i$ is defined as $\sum_{j=1}^n w_jX_j$ where $(w_1,\dots,w_n)\in\R_+^n$ is the exposure vector after risk sharing with $\sum_{j=1}^n w_j=a_i$, meaning that LRS preserves each agent's total exposure. LRS is commonly used in peer-to-peer insurance  \citep{feng2023peer}. Suitably chosen exposure vectors can simplify liability calculations, and reduce  variance for losses with finite second moments \citep{buhlmann1979optimal}. Introduced by \cite{denuit2012convex}, CMRS assigns agent $i$ the shared loss $m_i(S)=\mathbb E[a_iX_i\mid S]$ where $S=\sum_{i=1}^na_iX_i$ is the total loss in the pool. 
Main applications include allocating claims retained by peer-to-peer pools and distributing mortality credits in tontines \citep{denuit2019size,denuit2021risk}. Explicit solutions to $m_i(S)$ for dependent losses can be found in \cite{denuit2022conditional} and \cite{denuit2025no}. See also \cite{jiao2022axiomatic} and \cite{blier2026laplace} for  a  characterization  and a numerical route for the CMRS, respectively. Note that  CMRS remains well defined even if  the losses have infinite mean.

For finite-mean losses, classical results have shown that both LRS and CMRS can reduce the variability of losses. A commonly used tool for the analysis is convex order; a random loss is less variable if it is smaller in convex order. Under LRS, it is well-known that $\sum_{j=1}^n w_jX_j$ is less than $a_iX_i$ in convex order for exchangeable $X_1,\dots,X_n$ with finite mean; see, e.g., Property 3.4.48 of \cite{denuit2005actuarial} and Theorem 3.A.35 of \cite{SS07}. A relevant relation established by \cite{proschan1965peakedness} is that for iid symmetric log-concave risks $X_1,\dots,X_n$,  $\sum_{j=1}^n w_jX_j$ is less than $a_iX_i$ in peakedness order, i.e.,  $\sum_{j=1}^n w_jX_j$ is more concentrated. This result has been generalized to independent and non-identically distributed risks \citep{ma1998peakedness}, and finite-mean stable risks \citep{ibragimov2005new}. Both relations in convex order and peakedness order can be extended to the comparison of linear portfolios with arbitrary exposure vectors ranked by majorization order (see Definition \ref{def:majorization_order}). The benefit of risk sharing in convex order holds similarly for CMRS. For finite-mean losses $X_1,\dots,X_n$ with arbitrary dependence, \cite{denuit2012convex} have shown that $m_i(S)$ is smaller than $a_iX_i$ in convex order.  Additionally, if losses are independent, adding an extra  loss to the pool reduces each existing agent's risk position in convex order \citep{denuit2021risk,denuit2023risk}.


For infinite-mean losses, risk sharing can however incur a penalty. Under LRS, \cite{ibragimov2005new} has shown that more balanced exposure vectors reduce the peakedness for iid stable risks with tail parameter in $(0,1]$. Hence, LRS makes the loss portfolio more spread-out. If the stable risks are positive,  the peakedness order relation implies that risk sharing can produce stochastically larger losses, i.e.,
\begin{align}
\label{eq:main}
a_iX_i\le_{\rm st}\sum_{j=1}^n w_jX_j\quad\text{with}\quad\sum_{j=1}^n w_j=a_i,
\end{align} 
where $\le_{\rm st}$ denotes (usual) stochastic order (see Section \ref{sec:stochastic_order} for the definition). This relation for independent infinite-mean Pareto risks is first established by \cite{chen2025unexpected}, and it has been subsequently generalized to broader classes of loss models, incorporating more general marginal distributions and dependence structures  \citep{arab2025convex, muller2025some, chen2026risk}. \cite{chen2025diversification} and \cite{chen2026stochastic} have extended the comparison \eqref{eq:main} to portfolios with arbitrary exposure vectors using majorization order. In particular, \cite{chen2026stochastic} establish a more general class of loss distributions, called the inverted concave distributions, which includes the Pareto and Fr\'echet distributions with infinite mean. As noted by \cite{denuit2025comonotonicity},  CMRS reduces to LRS if $a_1=\dots=a_n$, i.e., $m_i(S)=S/n$,   and a diversification penalty for iid infinite-mean Pareto risks follows from \cite{chen2025unexpected}:
\begin{align}
\label{eq:main2}
a_iX_i\le_{\rm st}m_i( S).
\end{align}

Despite recent advances in understanding diversification penalties under LRS, most results focus on independent and negatively dependent risks. To the best of our knowledge, only \cite{chen2025diversification} have studied \eqref{eq:main} under the Pareto-Clayton model. Moreover, the effect of CMRS on infinite-mean losses remains under-examined in general settings. A natural question is whether relations \eqref{eq:main}, \eqref{eq:main2}, or even their stronger versions can hold for more general losses with positive dependence, which often appear in scenarios where a common shock can affect individual losses simultaneously. Examples of common shocks include natural disasters \citep{pagnottoni2022climate} and pandemics \citep{abuzayed2021systemic}.
Motivated by this, we employ background risk models for our study. Note that while \eqref{eq:main} can hold under general notions of negative dependence \citep{chen2025unexpected,chen2026risk}, the case of positive dependence is more subtle; such relation may not hold for positively quadrant dependent losses as discussed  in Remark \ref{positive_dependence_counterexample}.

The multiplicative background risk (MBR) model is a special case of a background risk model in which each loss takes the form $X_i=RY_i$. The positive common factor $R$ is independent of the  positive idiosyncratic risk vector $(Y_1,\dots,Y_n)$. MBR models are widely used in insurance, economics and finance \citep{furman2021multiplicative}, with applications including credit default risk measurement \citep{su2017form}, and the aggregation of dependent insurance losses \citep{semenikhine2018multiplicative}.  \cite{sarabia2016risk} derive explicit formulas to calculate risk measures Value-at-Risk and Expected Shortfall under the Pareto-Clayton model where each idiosyncratic risk $Y_i$ follows an exponential distribution, and the background risk $R$ follows an inverse gamma distribution; see \cite{sarabia2018aggregation} for other MBR models with idiosyncratic risks that are exponentially distributed. \cite{denuit2025no} obtain explicit solutions to the CMRS allocation for the Pareto-Clayton model. Flexible choices of both the background risk and idiosyncratic risks can accommodate heterogeneous risk exposures and heavy-tailed losses.


We summarize our contributions below.
\begin{itemize}
    \item We show in Theorem \ref{thm:common-multiplier} that several diversification penalties stronger than \eqref{eq:main} hold for MBR models with arbitrary idiosyncratic risks  if and only if the background risk has an inverted concave distribution. This characterization allows non-identically distributed and dependent idiosyncratic risks, and connects the concavity of Value-at-Risk to diversification penalties. 

    \item Theorem \ref{thm:radial-archimedean} establishes a diversification penalty under LRS for infinite-mean losses with $n$-monotone Archimedean generators. Theorem \ref{thm:Pareto-mixture} identifies the parameter conditions under which the diversification penalty holds for heavy-tailed Pareto risks coupled by a mixture of Archimedean copulas. Such mixtures allow dependence to vary across regimes while holding the Pareto marginals fixed, isolating the role of dependence in the diversification penalty.

    \item Theorem \ref{thm:inverted-concave-common-factor} shows that if the background risk in an MBR model is inverted concave, \eqref{eq:main2} holds.  This, to the best of our knowledge, is the first result for diversification penalty under CMRS in a general  setting.  For independent idiosyncratic risks that follow  stable distributions, Theorem \ref{thm: CMRS_stable} derives explicit CMRS allocations. The explicit solution to CMRS is used in Theorem \ref{thm: peakedness} to compare CMRS allocations with the initial losses in peakedness order.
\end{itemize}

The rest of the paper is organized as follows. Section \ref{sec:stochastic_order} briefly reviews stochastic orders, background risk models, and copulas. Section \ref{sec:LRS} establishes the diversification penalty under LRS for MBR models with an inverted concave background risk or inverted concave idiosyncratic risks. Section \ref{sec:Pareto} examines the diversification penalty under LRS for Pareto marginals coupled by mixtures of Archimedean copulas. Section \ref{sec:CMRS} develops a  diversification penalty under CMRS, and Section \ref{sec:CMRS_Stable} derives explicit CMRS allocations for stable risks and compares them with the original risks in peakedness order. Section \ref{sec:numerical} provides a short simulation study. Section \ref{sec:conclusion} concludes. The supplementary materials contain the proof of Theorem \ref{thm:Pareto-mixture}.

\subsection*{Notation and convention}
Denote the set of all positive integers by $\mathbb{N}$, the set of non-negative real numbers by $\mathbb{R}_{+}$, and the extended real line by $\overline R:=\R\cup\{-\infty,+\infty\}$. For $n\in\N$, denote by $[n]=\{1,\dots,n\}$. Throughout, we fix an integer $n\ge 2$. Define $\Delta_n=\{(\theta_1,\dots,\theta_n)\in [0,1]^n:  \sum_{i=1}^n \theta_i=1\}$. For $x,y \in \mathbb{R}$, write $x \wedge y = \min\{x,y\}$. For a random variable $X$, denote its cumulative distribution function, survival function, and density function by $F_X$, $\bar F_X=1-F_X$, and $f_X$, respectively. The essential infimum and essential supremum of $X$ are defined as $$\essinf X=\inf\{x\in\R: \p(X\le x)>0\}\mbox{   and   }\esssup X=\sup\{x\in\R: \p(X\le x)<1\}.$$ For $y \in \R$ with $\inf \varnothing=\infty$, let $f^{-1}$ be the left  inverse function of a function $f: \R\to \R$, defined as 
\begin{align*}
    f^{-1}(y)&=\inf\{x \in \R: f(x)\ge y\}. 
\end{align*} 
If two random variables $X$ and $Y$ have the same cumulative distribution function, we write $X\laweq Y$.

\section{Preliminaries}
\label{sec:stochastic_order}
This section reminds the readers of some well-known definitions and results of stochastic orders, background risk models, and copulas.

Stochastic order is used to compare and rank different probability distributions to determine which one is preferable under certain rational decision-making criteria \citep{hadar1969rules}. For two random variables $X$ and $Y$, $X$ is said to be smaller than $Y$ in  \emph{(usual) stochastic order}, denoted by $X \leq_{\rm st} Y$, if $\mathbb{P}(X \leq x) \geq \mathbb{P}(Y \leq x)$ for all $x \in \mathbb{R}$.  This means that if $X$ and $Y$ are treated as losses, any rational agent who prefers less loss will choose $X$ over $Y$. Flipping the sign, if $X$ and $Y$ represent profits, then any rational decision maker will choose $Y$ over $X$. For integrable  $X$ and $Y$, we say that $X$ is less than $Y$ in \emph{convex order}, denoted by $X \leq_{\rm cx} Y$, if $\mathbb{E}[\phi(X)] \le \mathbb{E}[\phi(Y)]$ for all convex functions $\phi:\mathbb{R} \to \mathbb{R}$ for which the two expectations are well defined. Convex order is used to compare the variability of random variables; e.g., if $X,Y$ have finite second moments, then $X\le_{\rm cx}Y$ implies $\mathrm{Var}(X)\le \mathrm{Var}(Y)$. For a new notion of convex order defined for random variables with possibly infinite mean, see \cite{CW26}. We refer to \cite{SS07} for the mathematics of various notions of stochastic orders.

A \emph{background risk model} $(X_1,\dots,X_n)$ is defined via the form
\begin{equation}
\label{eq:general-background risk}
 X_i=f(Y_i,R)\quad \text{for all}\quad i\in[n],
\end{equation}
where $R,Y_1,\dots,Y_n$ are positive random variables, and $f:\R_+^2\to\R_+$ is an increasing Borel measurable function with $f(0,r)=0$. Here, $R$ is interpreted as the \emph{background risk} and $(Y_1,\dots,Y_n)$ is a vector of \emph{idiosyncratic risks} independent of $R$. The conditions of $f$ ensure that a larger idiosyncratic loss or a more adverse background risk cannot reduce a loss. A particularly important background risk model is the \emph{multiplicative background risk (MBR) model}, given by
\begin{equation}
\label{eq:MBR}
 X_i=RY_i\quad \text{for all}\quad i\in[n].
\end{equation}
The class of MBR models includes the classical Pareto-Clayton model and various exponential mixture models, which are used in modelling insurance losses;  see Section \ref{sec:intro}. 

A \emph{copula} is a distribution function of a random vector with marginals uniformly distributed on $(0,1)$; see \cite{N06} for an introduction. By Sklar's Theorem \citep[Theorem 7.3]{MFE15}, for a random vector $(X_1,\dots,X_n)$ with marginal distributions $F_1,\dots,F_n$, there always exists a copula $C$ such that
\[
 \p(X_1\leq x_1,\dots,X_n\leq x_n)
 =C(F_1(x_1),\dots,F_n(x_n)), \quad (x_1,\dots,x_n)\in\R^n.
\]
The survival version of the Sklar's Theorem gives 
\[
 \p(X_1>x_1,\dots,X_n>x_n)
 =\bar C(\bar F_1(x_1),\dots,\bar F_n(x_n)),\quad (x_1,\dots,x_n)\in\R^n.
\]
where $\bar C$ is known as the \emph{survival copula}. Hence, a joint distribution function can always be decomposed into marginal distributions and a copula.  The connection of the MBR models with the class of Archimedean copulas is discussed in Section \ref{sec:archimedean_copula}.

\section{Linear risk sharing for MBR models}
\label{sec:LRS}
\subsection{Inverted concave background risk}
\label{sec: LRS_IC_background}
Suppose that there are $n$ agents in a risk sharing pool. Agent $i\in[n]$ has an initial loss $a_iX_i$, where $a_i>0$ and $(X_1,\dots,X_n)$ follows the MBR model \eqref{eq:MBR}. The total loss of the pool is defined as $S=\sum_{i=1}^n a_iX_i$. Denote by $A_i$ the loss position of agent $i\in[n]$ after risk sharing. The \emph{linear risk sharing (LRS)} \citep{buhlmann1979optimal} is defined by 
\[
 A_i=\sum_{j=1}^n a_{ij}X_j,\qquad \sum_{i=1}^n A_i=S, \qquad i\in[n],
\]
where $a_{ij}\geq0$ is agent $i$'s exposure to $X_j$ after risk sharing,  and $\sum_{j=1}^n a_{ij}=a_i$ for every $i\in[n]$, meaning that LRS preserves each agent's total exposure before and after risk sharing. We will investigate which exposure vector $(a_{i1},\dots,a_{in})$ can minimize the risk of $A_i$.


Majorization provides a useful way to compare the degree of diversification under LRS. For $(x_1,\dots,x_n)$, denote by $x_{[i]}$ the $i$-th largest component of $(x_1, \dots, x_n)$.
\begin{definition}
\label{def:majorization_order}
For $(\theta_1, \dots, \theta_n)$ and $(\eta_1, \dots, \eta_n)$ in $\R^n$, $(\theta_1, \dots, \theta_n)$ is said to be smaller than $(\eta_1, \dots, \eta_n)$ in \emph{majorization order}, denoted by $(\theta_1, \dots, \theta_n)\preceq(\eta_1, \dots, \eta_n)$ if
\[
\sum_{i=1}^n\theta_{i}=\sum_{i=1}^n\eta_{i}\quad \text{and}\quad\sum_{i=1}^k\theta_{[i]}\leq\sum_{i=1}^k\eta_{[i]}\quad\text{for all}\quad k=1,\dots,n-1.
\]
\end{definition}

Let $(\theta_1, \dots, \theta_n)$ and $(\eta_1, \dots, \eta_n)$ in $\R_+^n$ be exposure vectors. We say
 $(\theta_1, \dots, \theta_n)$ is more diversified  than $(\eta_1, \dots, \eta_n)$ if  $(\theta_1, \dots, \theta_n)\preceq(\eta_1, \dots, \eta_n)$; e.g., $$\left(\frac 1 n,\dots,\frac 1 n,\frac 1 n\right)\preceq\left(\frac{1}{n-1},\dots,\frac{1}{n-1},0\right)\preceq(1,\dots,0,0).$$  For exchangeable losses $Z_1,\dots,Z_n$ with finite mean, it is well-known that a more diversified portfolio is less risky in the sense of convex order, i.e., if $(\theta_1, \dots, \theta_n)\preceq(\eta_1, \dots, \eta_n)$, then
\(
\sum_{i=1}^n\theta_{i} Z_{i}\leq_{\mathrm{cx}} \sum_{i=1}^n\eta_{i} Z_{i}
\) \citep[Theorem 3.A.35]{SS07}.
Hence, more diversified portfolios are less variable in general. 

However, diversification may not be beneficial if losses have infinite mean.  \cite{chen2026stochastic} have shown that for iid losses $Z_1,\dots,Z_n$ that are inverted concave (see Definition \ref{def:IC}),  
\begin{equation}
\label{eq:dpl}
\eta_1Z_1+\dots+\eta_nZ_n \leq_{\st} \theta_1Z_1+\dots+\theta_nZ_n
\end{equation}
holds for any $(\theta_1, \dots, \theta_n)$ and $(\eta_1, \dots, \eta_n)$ in $\R_+^n$ with $(\theta_1, \dots, \theta_n)\preceq(\eta_1, \dots, \eta_n)$. Value-at-Risk (VaR) is one of the regulatory risk measures used in the insurance sector. The VaR of $X$ at level $p\in (0,1)$ is defined as the left  inverse function of $F$, that is,
\begin{equation*}
\VaR_{p}(X)=F^{-1}(p)=\inf\{x\in\R:F(x)\geq p\}.
\end{equation*} 
Note that $X\le_{\rm st}Y$ if and only if $\VaR_p(X)\le \VaR_p(Y)$ for all $p\in(0,1)$. Thus if $\eta_1=\sum_{i=1}^n\theta_i$, \eqref{eq:dpl} can be rewritten as 
\begin{equation}
\label{eq:dpl2}
\VaR_p(\theta_1Z_1+\dots+\theta_nZ_n)\ge \VaR_p(\theta_1Z_1)+\dots+\VaR_p(\theta_nZ_n)\quad\text{for all}\quad p\in(0,1).
\end{equation} 
Property \eqref{eq:dpl2}, after renormalizing the total exposure to 1, is introduced as the weighted universal VaR superadditivity by \cite{CLW26}, who also establish \eqref{eq:dpl2} for non-iid losses. We say that a loss vector $(Z_1,\dots,Z_n)$ satisfies the \emph{diversification penalty under LRS} if either \eqref{eq:dpl} or \eqref{eq:dpl2} holds. Clearly, both \eqref{eq:dpl} and \eqref{eq:dpl2} imply \eqref{eq:main}. We will show that the above diversification penalties hold in some general MBR models if and only if the background risk has an inverted concave distribution, as defined below.

\begin{definition}\label{def:IC}
Let $Z$ be a random variable with $\essinf Z=0$ and $\p(Z>0)=1$.\footnote{ \cite{chen2026stochastic} allow IC distributions to have a point mass at 0.} We say $Z$ or $F_Z$ is \emph{inverted concave}, denoted by $Z\in \mathrm{IC}$, if the function $x\mapsto \bar F_Z(1/x)$ is concave on $(0,\infty)$.
\end{definition}

Every $Z\in\mathrm{IC}$ has a continuous distribution function.  Note that the diversification penalty cannot hold for non-degenerate iid losses with finite mean; see Proposition 1(i) of \cite{chen2026stochastic}, and thus IC distributions have infinite mean. Examples of IC distributions include the Pareto and Fr\'echet distributions with infinite mean, which will be used as the primary examples of marginal distributions throughout this paper. See \cite{chen2026stochastic} for more examples.  A random variable $X$ follows the Pareto distribution with tail parameter $\alpha>0$, denoted by $X\sim {\rm Pareto}(\alpha)$, if its survival function is $$\mathbb{P}(X>x)=(1+x)^{-\alpha},\qquad x\geq0.$$ A random variable $Y$ follows the  Fr\'echet distribution with tail parameter $\beta>0$, denoted by $Y\sim {\rm \text{Fr\'echet}}(\beta)$, if its cumulative distribution function is $$\mathbb{P}(Y\leq y)=\exp(-y^{-\beta}),\qquad y\geq0.$$
If the tail parameters of the Pareto and the Fr\'echet distributions are less than or equal to 1, both distributions are IC and have infinite mean.

The following result characterizes the background risk $R$ that induces diversification penalties \eqref{eq:dpl} and \eqref{eq:dpl2} in MBR models. Random variables $Z_1,\dots,Z_n$ are exchangeable if $(Z_1,\dots,Z_n)\laweq(Z_{\sigma(1)},\dots,Z_{\sigma(n)})$ for every permutation $(\sigma(1),\dots,\sigma(n))$  of $(1,\dots,n)$.

\begin{theorem}
\label{thm:common-multiplier}
For $i\in[n]$, suppose that $X_i=RY_i$ where $R,Y_1,\dots,Y_n$ are positive random variables, and $R$ is independent of $(Y_1,\dots,Y_n)$. The following statements are equivalent.
\begin{enumerate}[(i)]
\item 
$R\in \mathrm{IC}$;
\item 
$R\laweq C(Z+1)$ where $Z\sim \mathrm{Pareto(1)}$ and $C$ is a positive random variable independent of $Z$;
\item 
For all $Y_1,\dots,Y_n$ and $p\in(0,1)$, $(w_1,\dots,w_n)\mapsto \VaR_p(w_1X_1+\dots+w_nX_n)$ is concave on $\R_+^n$;
\item 
For all $Y_1,\dots,Y_n$, $p\in(0,1)$, and  $(w_1,\dots,w_n)\in\R_+^n$, $$\VaR_p(w_1X_1+\dots+w_nX_n)\ge \VaR_p(w_1X_1)+\dots+\VaR_p(w_nX_n);$$ 
\item
For all exchangeable $Y_1,\dots,Y_n$, and $(\theta_1, \dots, \theta_n)$ and $(\eta_1, \dots, \eta_n)$ in $\R_+^n$ with  $(\theta_1, \dots, \theta_n)\preceq(\eta_1, \dots, \eta_n)$, $$\eta_1X_1+\dots+\eta_nX_n
 \leq_{\st} \theta_1X_1+\dots+\theta_nX_n.$$
\end{enumerate}
\end{theorem}

\begin{proof}
\underline{(i)$\Rightarrow$(ii)}:  Let $W=1/R$. For $x>0$, $\p(W\le x)=\p(R\ge 1/x)$.  As $R\in\mathrm{IC}$ and $R$ is continuous, the cumulative distribution function of $W$ is concave and thus unimodal on $(0,\infty)$. By Khintchine's characterization of unimodal distributions \cite[p.~158]{F71},  a distribution function is unimodal if and only if it is the distribution function of the product of two independent random variables of which one is uniformly distributed on $(0,1)$. Hence, there exist independent random variables $C>0$ and $U\sim \mathrm{Unif}(0,1)$ such that $R\laweq C/U$. Since $1/U-1\sim \mathrm{Pareto(1)}$, we have (ii).

\underline{(ii)$\Rightarrow$(iii)}: 
Write  $\bm{w}=(w_1,\dots,w_n)$. Let $S_{\bm{w}}=\sum_{i=1}^nw_iX_i$ and $Q_p(\bm{w})=\VaR_p(\sum_{i=1}^nw_iX_i)$. By (ii), we can take $R=C/U$ where $C$ is independent of $U\sim \mathrm{Unif}(0,1)$ and $(Y_1,\dots,Y_n)$.  Hence, $S_{\bm{w}}=B_{\bm{w}}/U$ where $B_{\bm{w}}=C\sum_{i=1}^n w_iY_i$. For $s>0$, by independence of $U$ and $B_{\bm w}$, we have
\[
\mathbb P(S_{\bm{w}}>s)=\E\!\left[\p (S_{\bm{w}}>s \mid B_{\bm{w}})\right]=\E\!\left[\p(U<B_{\bm{w}}/s \mid B_{\bm{w}})\right]=\E\!\left[\min\!\left\{\frac{B_{\bm{w}}}s,1\right\}\right].
\]
Therefore, $\bm{w}\mapsto \p(S_{\bm{w}}>s)$ is concave on $ \R_+^n$. For $\bm{\theta}, \bm{\eta}\in \R_+^n$ and $\lambda\in[0,1]$, we have
\[
\frac{\lambda\bm{\theta}+(1-\lambda)\bm{\eta}}{\lambda Q_p(\bm{\theta})+(1-\lambda)Q_p(\bm{\eta})}=\frac{\lambda Q_p(\bm{\theta})}{\lambda Q_p(\bm{\theta})+(1-\lambda)Q_p(\bm{\eta})}\frac{\bm{\theta}}{Q_p(\bm{\theta})}+\frac{(1-\lambda)Q_p(\bm{\eta})}{\lambda Q_p(\bm{\theta})+(1-\lambda)Q_p(\bm{\eta})}\frac{\bm{\eta}}{Q_p(\bm{\eta})}.
\]
Define 
$$A=\frac{\lambda\bm{\theta}+(1-\lambda)\bm{\eta}}{\lambda Q_p(\bm{\theta})+(1-\lambda)Q_p(\bm{\eta})}\mbox{   and   }B=\frac{\lambda Q_p(\bm{\theta})}{\lambda Q_p(\bm{\theta})+(1-\lambda)Q_p(\bm{\eta})}.$$
By concavity of the survival  function of $S_{\bm w}$, we have
\begin{align*}
\p\left(S_{\lambda\bm{\theta}+(1-\lambda)\bm{\eta}}>\lambda Q_p(\bm{\theta})+(1-\lambda)Q_p(\bm{\eta})\right)&=\p\left(S_{A}>1\right)\\
&\ge B\p\left(S_{\bm{\theta}/Q_p(\bm{\theta})}>1\right)+(1-B)\p\left(S_{\bm \eta/Q_p(\bm \eta)}>1\right)\\
&=B\p\left(S_{\bm{\theta}}>Q_p(\bm{\theta})\right)+(1-B)\p\left(S_{\bm \eta}> Q_p(\bm \eta)\right)=1-p.
\end{align*}
The last equality follows from the continuity of $S_{\bm \theta}$ and $S_{\bm \eta}$ as $R$ is continuous. By the definition of quantile,
\[
Q_p(\lambda\bm{\theta}+(1-\lambda)\bm{\eta}) \ge\lambda Q_p(\bm{\theta})+(1-\lambda)Q_p(\bm{\eta}), \quad 0\le\lambda\le1.
\]
Thus, we have (iii).

\underline{(iii)$\Rightarrow$(iv)}: Denote by $\mathbf{e}_i$  the vector whose $i$-th coordinate is one and whose other coordinates are zero. Note that VaR is positive homogeneous. For $\bm{w}=(w_1,\dots,w_n)\in\R_+^n$, we have $\bm{w}=\sum_{i=1}^n w_i \mathbf{e}_i$. By (iii), 
\begin{align*}
\VaR_p(w_1X_1+\dots+w_nX_n)=Q_p(\bm w)
&=\left(\sum_{i=1}^nw_i\right)Q_p\left(\frac{\sum_{i=1}^nw_i\mathbf{e}_i}{\sum_{i=1}^nw_i}\right)\\
&\ge \left(\sum_{i=1}^nw_i\right) \sum_{i=1}^n \frac{w_i}{\sum_{i=1}^nw_i} Q_p(\mathbf{e}_i)\\
&= \sum_{i=1}^n w_i Q_p(\mathbf{e}_i)=\sum_{i=1}^n\VaR_p(w_iX_i).
\end{align*}

\underline{(iii)$\Rightarrow$(v)}: A doubly stochastic matrix is a square matrix of non-negative real numbers, whose row and column sums are all 1. Write $\bm \theta=(\theta_1,\dots,\theta_n)$ and $\bm \eta=(\eta_1,\dots,\eta_n)$. Denote by $\mathcal M_n$ the set of $n\times n$ doubly stochastic matrices. By the equivalent characterizations of majorization order \citep[Section 1.A.3]{MOA11}, $\bm\theta\preceq\bm\eta$ if and only if there exists $\Lambda \in \mathcal M_n$ such that $\bm\theta=\Lambda \bm\eta$.  By Birkhoff's Theorem \citep[Theorem 2.A.2]{MOA11}, the set $\mathcal M_n$ of doubly stochastic matrices is the convex hull of permutation matrices, i.e., $\Pi \bm\theta$ gives a permutation of $\bm\theta$, where $\Pi\in\mathcal M$. Let $\Pi_1,\dots,\Pi_{n!}$ be all different $n$-permutation matrices. Thus, for any $\Lambda \in \mathcal M_n$,  there exists  $(\lambda_1,\dots,\lambda_{n!})\in \Delta_{n!}$ such that
$
\Lambda= \sum_{k=1}^{n!}\lambda_k \Pi_k.
$
Therefore, 
\begin{align*}
Q_p(\bm\theta)=Q_p\left(\sum_{k=1}^{n!}\lambda_k \Pi_k\bm\eta\right)\ge \sum_{k=1}^{n!}\lambda_kQ_p\left( \Pi_k\bm\eta\right)=\sum_{k=1}^{n!}\lambda_kQ_p\left( \bm\eta\right)=Q_p\left( \bm\eta\right).
\end{align*}
The second last equality is due to the exchangeability of $X_1,\dots,X_n$. Then use the equivalence that for any random variables $X$ and $Y$, $\VaR_p(X)\le \VaR_p(Y)$ for all $p\in(0,1)$ if and only if $X\le_{\rm st}Y$.

\underline{(iv) or (v)$\Rightarrow$(i)}: Fix $a,b>0$ and let $Y_1,Y_2$ be independent and  take values $a$ and $b$ with equal probabilities.  Applying (iv) or (v) at the threshold $x>0$ gives
\[
\frac12\bar F_R\left(\frac{x}{a}\right)
 +\frac12\bar F_R\left(\frac{x}{b}\right)
 \leq
 \frac14\bar F_R\left(\frac{x}{a}\right)
 +\frac12\bar F_R\left(\frac{2x}{a+b}\right)
 +\frac14\bar F_R\left(\frac{x}{b}\right).
\]
Consequently,
\[
 \bar F_R\left(\frac{2x}{a+b}\right)
 \geq
 \frac12\bar F_R\left(\frac{x}{a}\right)
 +\frac12\bar F_R\left(\frac{x}{b}\right).
\]
As $a$, $b$, and $x$ are arbitrary, the function $u\mapsto\bar F_R(1/u)$ is midpoint concave on $(0,\infty)$.  It is bounded and hence concave. This proves (i).
\end{proof}


Without imposing any restrictions on the marginal distributions or dependence structure of the positive idiosyncratic risks, Theorem \ref{thm:common-multiplier} (iii) shows that combining portfolios produces a VaR no smaller than the corresponding weighted average of the  VaRs of portfolios at any confidence level $p\in(0,1)$; it will also be used to establish a diversification penalty for conditional mean risk sharing in MBR models (Theorem \ref{thm:inverted-concave-common-factor}). This leads to the weighted universal VaR superadditivity in (iv), which implies that the smallest VaR of the portfolio can be achieved by putting all exposures on a single loss that gives the smallest individual VaR \citep[Proposition 4]{CLW26}. Under exchangeability of the  idiosyncratic risks, (v) establishes the diversification penalty \eqref{eq:dpl}. Moreover, (ii) shows that the admissible background risk is a scale mixture of the shifted Pareto$(1)$ random variable.

\begin{example}
\label{ex:PC}
Let $\mathcal B$ be a collection of non-empty subsets of $[n]$ such that $\bigcup_{I\in\mathcal B}I=[n]$. For each non-empty $I\subseteq[n]$, let $E_I$ be independent exponential random variables with rates $\lambda_I>0$. For $i\in[n]$, suppose that $$X_i=RY_i \quad \text{where}\quad Y_i=\min\{E_I:I\in\mathcal B ~\text{and}~ i\in I\}\quad \text{and}\quad R^{-1}\sim\operatorname{Gamma}(\alpha,1),$$ where $R$ is independent of $(Y_1,\dots,Y_n)$ and $0<\alpha \leq 1$. Here, $Y_1,\dots,Y_n$ can be non-identically distributed, dependent, and may not be idiosyncratic as some of them may be affected by common exponential random variables. This construction corresponds to Equation 6.2.3 of \cite{A15} and has been used by \cite{MF26} to develop a new class of multivariate models with Pareto marginals. Since $R\in\mathrm{IC}$ \citep[Example 3]{chen2026stochastic}, Theorem \ref{thm:common-multiplier} (iv) gives the weighted universal VaR superadditivity \eqref{eq:dpl2}.

Taking $\mathcal{B}=\{\{1\},\dots,\{n\}\}$ and $\lambda_I=1$ for all $I\in\mathcal B$ gives the Pareto-Clayton($\alpha$) model: for $i\in[n]$, 
\begin{equation}
\label{eq:PC}
R^{-1}\sim\operatorname{Gamma}(\alpha,1) \quad \text{and} \quad Y_i\overset{\mathrm{iid}}{\sim}\operatorname{Exp}(1),
\end{equation}
where $0<\alpha\le1$. For $x_1,\dots,x_n>0$, the joint survival function of $(X_1,\dots,X_n)$ is given by
$
 \p(X_1>x_1,\dots,X_n>x_n)
 =\left(1+\sum_{i=1}^n x_i \right)^{-\alpha}.
$
Hence, $X_i\sim \mathrm{Pareto}(\alpha)$ for $i\in[n]$.  The survival copula of $(X_1,\dots,X_n)$ is the  Clayton copula with parameter $\theta=1/\alpha \geq 1$, given by
\begin{equation*}
    \bar C(u_1,\dots,u_n)
 =\left(\sum_{i=1}^n u_i^{-\theta}-n+1\right)^{-1/\theta},\quad (u_1,\dots,u_n)\in [0,1]^n.
\end{equation*}
Since $R$ is IC , and $Y_1,\dots,Y_n$ are exchangeable, Theorem \ref{thm:common-multiplier} establishes  \eqref{eq:dpl} for  the Pareto-Clayton model, which is first studied in Theorem 3 of \cite{chen2025diversification}.
\end{example}

\begin{figure}[htbp]
\centering
\includegraphics[width=1\textwidth]{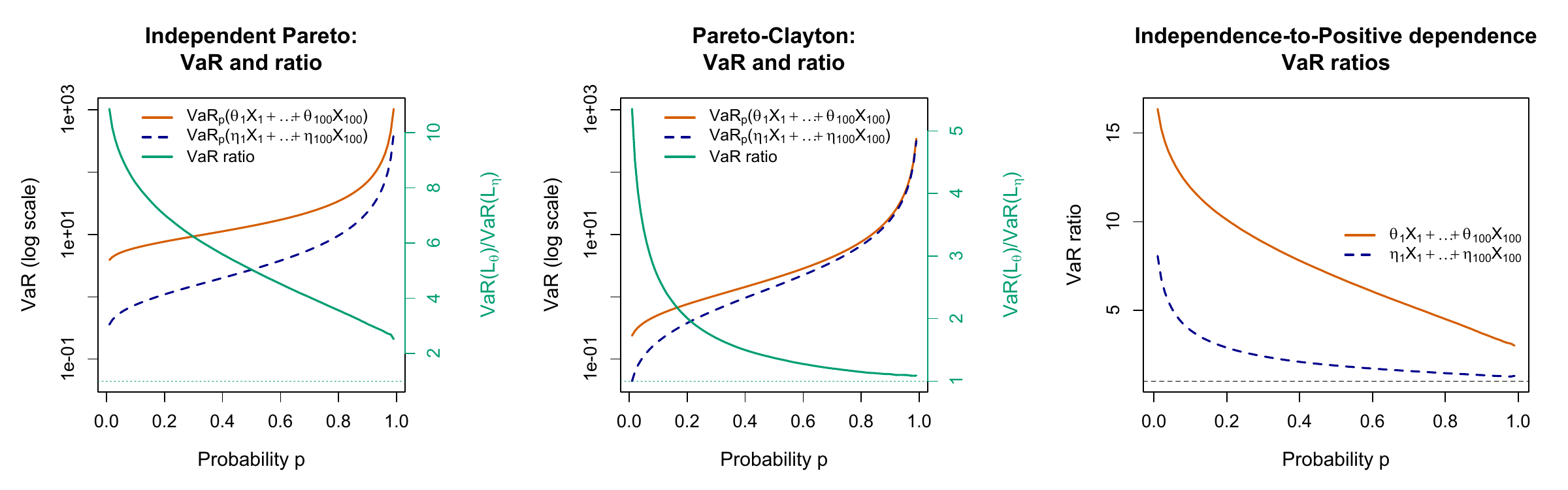}
\captionsetup{font=small}
\caption{Simulated $\VaR_p$ for $\theta_1X_1+\dots+\theta_{100}X_{100}$ and $\eta_1X_1+\dots+\eta_{100}X_{100}$ for independent $\mathrm{\text{Pareto}}(0.8)$ (left), Pareto-Clayton($0.8$) model (middle), and independence-to-positive dependence VaR ratio (right) for $p\in(0,1)$.}
\label{fig:pareto_clayton}
\end{figure}

Figure \ref{fig:pareto_clayton} shows the diversification penalty for $n=100$ with Pareto$(0.8)$ marginals under independence and Pareto-Clayton$(0.8)$ dependence by simulations. At every probability $p\in(0,1)$, the equally weighted portfolio with $(\theta_1,\dots,\theta_{100})=(0.01, \dots,0.01)$ has a higher estimated VaR than the less diversified portfolio with $(\eta_1,\dots,\eta_{100})=(0.95,0.05/99,\dots,0.05/99)$ in both models. For each portfolio, the empirical VaR is higher under independence than under Pareto-Clayton  dependence at every plotted probability level $p$.

\begin{remark}[General positive dependence]
\label{positive_dependence_counterexample}
As losses are more likely to occur together under positive dependence than under independence, a common belief is that diversification can lead to more penalties when transitioning from independence to positive dependence. We will show that this is not generally true.

Suppose that $X_1,X_2\sim F$ are iid positive and non-degenerate risks such that for  $(\theta_1,\theta_2)\in\Delta_2$, $X_1 \leq_{\mathrm{st}} \theta_1X_1+\theta_2X_2$. We construct positively dependent $Y_1, Y_2\sim F$ such that $Y_1 \not\leq_{\mathrm{st}} \theta_1Y_1+\theta_2Y_2$. Fix $p\in(0,1)$, and let $A$ be an event with $\p(A)=p$. Additionally, let $U_1,U_2 \overset{\mathrm{iid}}{\sim}\mathrm{Unif}(0,p)$, and $U_3,U_4 \overset{\mathrm{iid}}{\sim}\mathrm{Unif}(p,1)$, where the random vector $(U_1,U_2,U_3,U_4)$ is independent of $A$. Define the pair
\[
U = U_1\mathds{1}_{A} + U_3\mathds{1}_{A^c}
\qquad \mbox{and}\qquad
V = U_2\mathds{1}_{A} + U_4\mathds{1}_{A^c}.
\]
Then $U,V\sim \mathrm{Unif}(0,1)$. Define  $Y_1=F^{-1}(U)$ and $Y_2=F^{-1}(V)$. We have $Y_1,Y_2\sim F$ \citep[Proposition 7.2]{MFE15}. For $(u_1, u_2) \in [0,1]^2$, the copula of $(U,V)$ is
\[
C(u_1,u_2)
=
\begin{cases}
\frac{u_1u_2}{p},
&0\le u_1,u_2\le p,\\[8pt]
p+\frac{(u_1-p)(u_2-p)}{1-p},
&p<u_1,u_2\le1,\\[8pt]
\min(u_1,u_2),
&\text{otherwise}.
\end{cases}
\] 
Therefore, $C(u_1,u_2)\ge u_1u_2$ for all $(u_1,u_2)\in[0,1]^2$, and $U,V$ are positively quadrant dependent \citep[Definition 5.2.1]{N06} and so are $Y_1,Y_2$.  For $x<\theta_1 \VaR_p(Y_1)+\theta_2 \VaR_p(Y_2)$,  we have
$
\p(\theta_1Y_1+\theta_2Y_2\le x)=\p(\theta_1Y_1+\theta_2Y_2\le x \mid A)\p(A).
$
Conditional on $A$, since $Y_1,Y_2$ have finite mean, are iid, bounded and non-degenerate, $\p(\theta_1Y_1+\theta_2Y_2\le x \mid A)\le \p(Y_1\le x \mid A)$ cannot hold for all $x<\theta_1 \VaR_p(Y_1)+\theta_2 \VaR_p(Y_2)$. Hence, $Y_1 \not\leq_{\mathrm{st}} \theta_1Y_1+\theta_2Y_2$.
\end{remark}

\begin{figure}[htbp]
\centering
\includegraphics[width=1\textwidth]{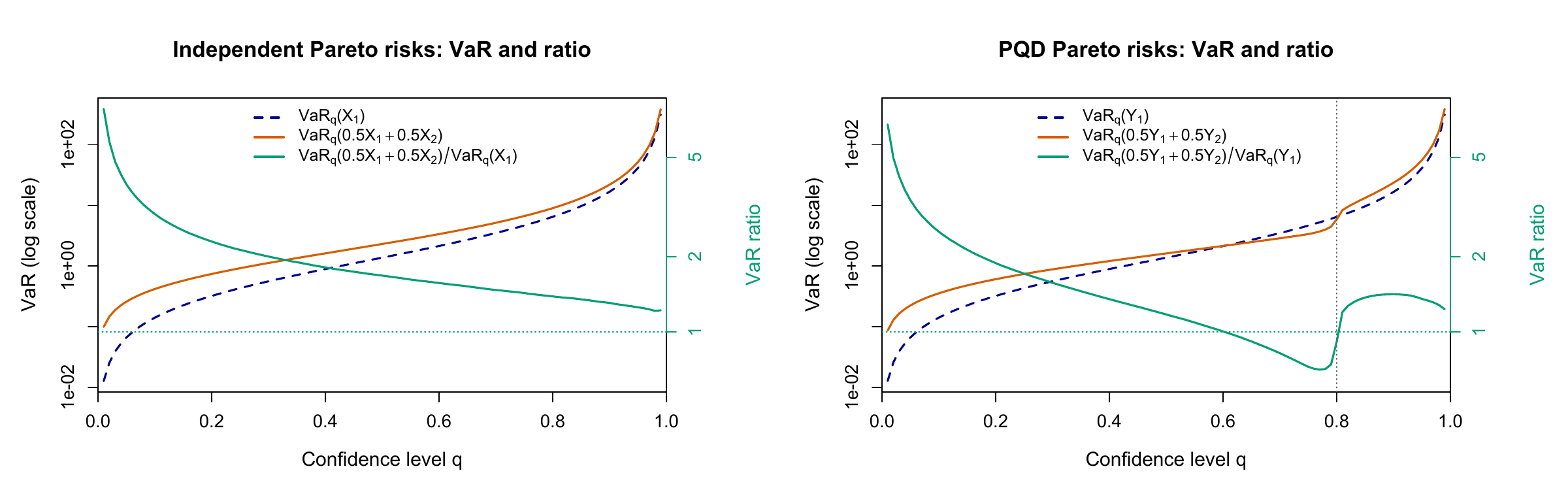}
\captionsetup{font=small}
\caption{Simulated $\VaR_q$ for $\mathrm{Pareto}(0.8)$ risks under independence (left), and the positive dependence specified in Remark \ref{positive_dependence_counterexample} (right) for $q\in(0,1)$.}
\label{fig:pareto_pqd}
\end{figure}

As an illustration, Figure \ref{fig:pareto_pqd} plots $\VaR_q(X_1), \VaR_q(0.5X_1+0.5X_2)$ and $\VaR_q(0.5X_1+0.5X_2)/\VaR_q(X_1)$ under independence, and $\VaR_q(Y_1), \VaR_q(0.5Y_1+0.5Y_2)$ and $\VaR_q(0.5Y_1+0.5Y_2)/\VaR_q(Y_1)$ under positive quadrant
dependence for $q\in(0,1)$ by simulations, where $X_1,X_2,Y_1,Y_2\sim \mathrm{Pareto}(0.8)$. At the chosen $\p(A)=p=0.8$, $\VaR_q(0.5Y_1+0.5Y_2)$ can have lower or higher VaR than $\VaR_q(Y_1)$ depending on the confidence level $q$, showing how the diversification penalty can fail.

\subsection{Inverted concave idiosyncratic risks}
\label{sec:archimedean_copula}
In Theorem \ref{thm:common-multiplier}, we have shown the diversification penalty when the background risk is IC. In this section, we consider the diversification penalty under the case where the (shifted) idiosyncratic risks are IC.

\begin{lemma}
\label{thm:conditional-dp}
For $i\in[n]$, suppose that $X_i=f(Y_i,R)$ where $R, Y_1,\dots,Y_n$ are positive random variables,  $Y_1,\dots,Y_n$ are iid conditional on $R$, and $f:\R_+^2\to\R_+$ is an increasing Borel measurable function. If $f(Y_i,r)\in\mathrm{IC}$ for every $r$ in the support of $R$,
 for any $(\theta_1, \dots, \theta_n)$ and $(\eta_1, \dots, \eta_n)$ in $\R_+^n$ with $(\theta_1, \dots, \theta_n)\preceq(\eta_1, \dots, \eta_n)$, $\eta_1X_1+\dots+\eta_nX_n
 \leq_{\st}
 \theta_1X_1+\dots+\theta_nX_n$.
\end{lemma}

\begin{proof}
For a fixed $r>0$ and $i\in[n]$, define $\tilde X_i=f(Y_i,r)$. These variables are iid and belong to $\mathrm{IC}$. Hence, for $(\theta_1, \dots, \theta_n)\preceq(\eta_1, \dots, \eta_n)$, we have $\eta_1\tilde{X}_1+\dots+\eta_n\tilde{X}_n \leq_{\st} \theta_1\tilde{X}_1+\dots+\theta_n\tilde{X}_n$ \citep[Theorem 1]{chen2026stochastic}. Since  usual stochastic order is preserved under a common mixture \citep[Theorem 1.A.3(d)]{SS07}, the desired unconditional stochastic relation is achieved.
\end{proof}

This lemma yields a conditional route to the diversification penalty. Together with Theorem \ref{thm:common-multiplier}, it   gives the following proposition: in MBR models with iid idiosyncratic risks, if the idiosyncratic risks or the background risk is IC, then the diversification penalty \eqref{eq:dpl} holds. In Section \ref{sec:Pareto}, we further use Lemma \ref{thm:conditional-dp} to obtain the diversification penalty for Pareto marginals coupled with various Archimedean copulas; these models are not MBR models in general.

\begin{proposition}
\label{prop:two-routes}
For $i\in[n]$, suppose that $X_i=RY_i$ where $Y_1,\dots,Y_n$ are iid positive random variables, and $R>0$ is independent of $(Y_1,\dots,Y_n)$. Let $(\theta_1,\dots,\theta_n)$ and $(\eta_1,\dots,\eta_n)$ in $\R_+^n$, and $(\theta_1, \dots, \theta_n)\preceq(\eta_1, \dots, \eta_n)$. 
\begin{enumerate}
\item[(i)] If $R\in\mathrm{IC}$, then $\eta_1X_1+\dots+\eta_nX_n \leq_{\st} \theta_1X_1+\dots+\theta_nX_n$.
\item[(ii)] If $Y_i\in\mathrm{IC}$, then $\eta_1X_1+\dots+\eta_nX_n \leq_{\st} \theta_1X_1+\dots+\theta_nX_n$.
\end{enumerate}
\end{proposition}


In the following, we will show that the diversification penalty \eqref{eq:dpl} holds under the MBR models with shifted IC idiosyncratic risks and Archimedean copulas. We first give the definitions of both the $n$-monotone and completely monotone Archimedean copulas. An \emph{Archimedean copula} satisfies the following form  
\begin{equation}
\label{eq:archimedean_copula}
C(u_1,\dots,u_n)
=\psi\left(\sum_{i=1}^n\psi^{-1}(u_i)\right),\quad (u_1,\dots,u_n)\in [0,1]^n,
\end{equation}
where $\psi:[0,\infty)\to[0,1]$ is a decreasing and continuous function that satisfies  $\psi(0)=1$ and $\lim_{t\to\infty}\psi(t)=0$. The function $\psi$ is known as the generator of the copula \citep{N06}.

Let $a,b\in\overline{\mathbb{R}}$ with $a<b$, and let $n\geq 2$ be an integer. A real function $f$ is called \emph{$n$-monotone} on $(a,b)$ if it is differentiable up to order $n-2$ and for any $x\in(a,b)$, the derivatives satisfy $(-1)^k f^{(k)}(x)\geq 0$ for all $k=0,1,\dots,n-2$, where $f^{(0)}=f$, and further if $(-1)^{n-2}f^{(n-2)}$ is non-increasing and convex on $(a,b)$. The function $f$ is called \emph{completely monotone} on $(a,b)$ if it has derivatives of all orders on $(a,b)$ and $(-1)^k f^{(k)}(x)\geq 0$ for any $x\in(a,b)$ \citep[Definition 2.3]{mcneil2009multivariate}. Thus, complete monotonicity implies $n$-monotonicity. For a fixed dimension $n$, the Archimedean copula defined in \eqref{eq:archimedean_copula} is an $n$-monotone (resp.~completely monotone) copula if and only if the generator $\psi$ is $n$-monotone (resp.~completely monotone) on $[0,\infty)$ \citep[Theorem 2.2]{mcneil2009multivariate}. 

Consider the MBR model defined as
\begin{equation*}
\label{eq:reciprocal-radial}
X_i=RY_i=\frac{R}{D_i},\qquad i\in[n],
\end{equation*}
where $(D_1,\dots,D_n)\sim\operatorname{Dirichlet}(1,\dots,1)$ has a uniform distribution on the $n$-dimensional simplex, and is independent of $R>0$. By Definition 3.1 and Theorem 3.1(ii) of \cite{mcneil2009multivariate}, $(X_1,\dots,X_n)$ has an $n$-monotone Archimedean copula.


The marginal tail behaviour is explicit with $D_i\sim{\rm Beta}(1,n-1)$. Let $Y_i=1/D_i$ for $i\in[n]$, it satisfies
\[
    \p(Y_i-1\leq y)
    =\int_{1/(y+1)}^{1}(n-1)(1-s)^{n-2}\,ds
    =\left(1-\frac{1}{y+1}\right)^{n-1},
    \qquad y\geq0.
\]
The distribution of $Y_i-1$ is known as the Stoppa distribution and it is IC by Table 1 of \cite{chen2026stochastic}.

\begin{theorem}
\label{thm:radial-archimedean}
For $i\in[n]$, suppose that $X_i=R/D_i$ where $(D_1,\dots,D_n)\sim\operatorname{Dirichlet}(1,\dots,1)$ is independent of $R>0$. For any $(\theta_1, \dots, \theta_n)$ and $(\eta_1, \dots, \eta_n)$ in $\R_+^n$ with $(\theta_1, \dots, \theta_n)\preceq(\eta_1, \dots, \eta_n)$, $$\eta_1X_1+\dots+\eta_nX_n
 \leq_{\st}
 \theta_1X_1+\dots+\theta_nX_n.$$
\end{theorem}

\begin{proof}
Since $R$ is independent of $(D_1,\dots,D_n)$, it suffices to prove that $\sum_{i=1}^n \eta_i/D_i \leq_{\st} \sum_{i=1}^n \theta_i/D_i$ \citep[Theorem 1.A.3(b)]{SS07}. Since $D_1,\dots,D_n$ is exchangeable, permuting the weights does not change the distribution of $\sum_{i=1}^n w_i/D_i$. Moreover, $(\theta_1, \dots, \theta_n)\preceq(\eta_1, \dots, \eta_n)$ implies that $(\theta_1, \dots, \theta_n)$ can be obtained from $(\eta_1, \dots, \eta_n)$ by finitely many $T$-transforms \citep[][Section 1.A.3]{MOA11}; $(\theta_1, \dots, \theta_n)$ is a $T$-transform of $(\eta_1, \dots, \eta_n)$ if for some $\lambda \in [0,1]$ and $i,j\in [n]$ with $i \neq j$,
\begin{equation}
\label{eq:t-tran}
\theta_i=\lambda \eta_i + (1-\lambda)\eta_j,~\theta_j= (1-\lambda)\eta_i+\lambda \eta_j,
\end{equation}
and $\theta_k=\eta_k$ for $k\in [n]\setminus \{i,j\}$. Suppose that \eqref{eq:t-tran} holds with $\theta_k, \eta_k$ fixed for all $k\neq i,j$. It therefore suffices to prove that $\eta_i/D_i + \eta_j/D_j\leq_{\st}  \theta_i/D_i+\theta_j/D_j$ given $D_k$, $k\neq i,j$. The case $\theta_i+\theta_j=0$ is trivial. We assume that $\theta_i+\theta_j>0$.

Conditional on $D_k=d_k$ for $k\neq i,j$,  $D_i+D_j=t$ where $t=1-\sum_{k\notin\{i,j\}}d_k>0$.  Thus $D_i$ can take any value on $(0,t)$, with $D_j=t-D_i$. Let $(D_i^*, D_j^*)$ follow the same distribution of $(D_i, D_j)$ conditional on $ D_k=d_k$ for $k\neq i,j$. Since the Dirichlet vector $(D_1,\dots, D_n)$ is uniform on the simplex, the density of $D_i^* $ is constant on $(0,t)$, and equals $1/t$ \citep[Theorem 2.5]{ng2011dirichlet}. Hence, we have $U=D_i^*/t\sim\operatorname{Unif}(0,1)$, 
\[
D_i^*=tU,\quad\text{and} \quad D_j^*=t(1-U).
\]

Let  $a,b>0$ with $a+b=\theta_i+\theta_j>0$. Define 
\[
h_{a,b}(u)=\frac{a}{u}+\frac{b}{1-u}, \qquad 0<u<1.
\]
For $y>0$, solving the inequality of $h_{a,b}(u)\leq y$ is equivalent to solving the quadratic inequality of $yu^2-(y+a-b)u+a\leq0$. The interval that satisfies the inequality is non-empty if $y\geq(\sqrt a+\sqrt b)^2$ or  $y\leq(\sqrt a-\sqrt b)^2$. If $y\geq(\sqrt a+\sqrt b)^2$, the roots of $yu^2-(y+a-b)u+a\leq0$ are in $(0,1)$. The roots lie outside of $(0,1)$ if $y<(\sqrt a-\sqrt b)^2$.  Then, we have
\[
   \p\left(\frac{a}{D_i^*}+\frac{b}{D_j^*}\le \frac yt\right)= \p(h_{a,b}(U)\leq y)=
    \begin{cases}
        0, & y<(\sqrt a+\sqrt b)^2,\\[3pt]
        \displaystyle
        L_y(a,b)=\frac{\sqrt{(y-a-b)^2-4ab}}{y},
        & y\geq(\sqrt a+\sqrt b)^2.
    \end{cases}
\]
The function $L_y(a,b)$ is Schur-convex in  $(a,b)$ and $(a,b)\mapsto(\sqrt a+\sqrt b)^2$ is Schur-concave on $\R_+^2$. Since $(\theta_1, \theta_2)\preceq(\eta_1, \eta_2)$, for every $x>0$, we have
\[
\p\left(
    \frac{\theta_1}{D_i^*}+\frac{\theta_2}{D_j^*}\leq x \right) \leq \p\left(
    \frac{\eta_1}{D_i^*}+\frac{\eta_2}{D_j^*}\leq x
\right).
\]
When one of $\theta_i,\theta_j$ is 0, as $D_i^*,D_j^*$ are continuous and $D_i^*+D_j^*=t$, the above inequality can be shown by using a limiting argument. The proof is complete.
\end{proof}

Many important examples of the $n$-monotone Archimedean copulas come from the subclass of  completely monotone Archimedean copulas.  Consider the MBR model defined as 
\begin{equation}
\label{eq:archimedean-representation}
X_i=RY_i=\frac{R}{E_i},\qquad i\in[n]
\end{equation}
where $E_1,\dots,E_n$ are independent exponential random variables with rate 1, independent of $R>0$, and $R$ satisfies  $\psi(t)=\E[e^{-tR}]$ for some completely monotone function $\psi$.
It gives the Marshall-Olkin frailty construction \citep{marshall1988families}. For $x>0$, the marginal distribution is determined by
\[
F_{X_i}(x)=\mathbb{E}\!\left[\mathbb{P}\!\left(E_i\geq \frac{R}{x}\,\middle|\,R\right)\right]=\mathbb{E}\!\left[e^{-R/x}\right]=\psi(1/x).
\] The copula of $(X_1,\dots,X_n)$ is a completely monotone Archimedean copula with generator $\psi$ \citep[Proposition 7.51]{MFE15}. The idiosyncratic risk $Y_i=1/E_i \sim  \mathrm{\text{Fr\'echet}}(1)$ and is thus IC. By Proposition \ref{prop:two-routes} (ii), we have the diversification penalty \eqref{eq:dpl}. This is also implied by Theorem \ref{thm:radial-archimedean}; we can rewrite \eqref{eq:archimedean-representation} as
$$X_i=\frac{R}{E_i}=\frac{R/\sum_{i=1}^nE_i}{E_i/\sum_{i=1}^nE_i},$$
where
$$\left(\frac{E_1}{\sum_{i=1}^nE_i},\dots,\frac{E_n}{\sum_{i=1}^nE_i}\right)\laweq (D_1,\dots,D_n)$$
is independent of $\sum_{i=1}^nE_i$, by the Gamma-Dirichlet decomposition \citep[proof of Theorem 2.1]{ng2011dirichlet}.
\begin{example}[Fr\'echet-Gumbel($\delta$) model]
\label{ex:FG}
For $i\in[n]$, let
\begin{equation}
\label{eq:FG}
X_i=RY_i,\quad
R\sim\operatorname{Positive\ Stable}(\delta),\quad
Y_i\overset{\mathrm{iid}}{\sim}\text{Fr\'echet}(1),
\end{equation}
where $R$ is independent of $(Y_1,\dots,Y_n)$, and $R$ has the Laplace transform of $\mathbb{E}[e^{-tR}]=e^{-t^\delta}$ for $t\ge0$ and $0<\delta\le1$ (see \eqref{eq:stable-definition} for the definition of stable distributions). For $x_1,\dots,x_n>0$, the joint distribution function of $(X_1,\dots, X_n)$ is given by $\mathbb{P}(X_1\le x_1,\dots,X_n\le x_n)=\exp(-(\sum_{i=1}^nx_i^{-1})^\delta)$. Hence, $X_i\sim\mathrm{\text{Fr\'echet}}(\delta)$ for $i\in[n]$. The copula of $(X_1,\dots, X_n)$ is the Gumbel copula with parameter $\theta=1/\delta \geq 1$, given by
\[
C(u_1,\dots,u_n)=\exp\left\{-\left(\sum_{i=1}^n(-\log u_i)^\theta\right)^{1/\theta}\right\}, \quad (u_1,\dots,u_n)\in [0,1]^n.
\]
\end{example}

\begin{figure}[htbp]
\centering
\includegraphics[width=1\textwidth]{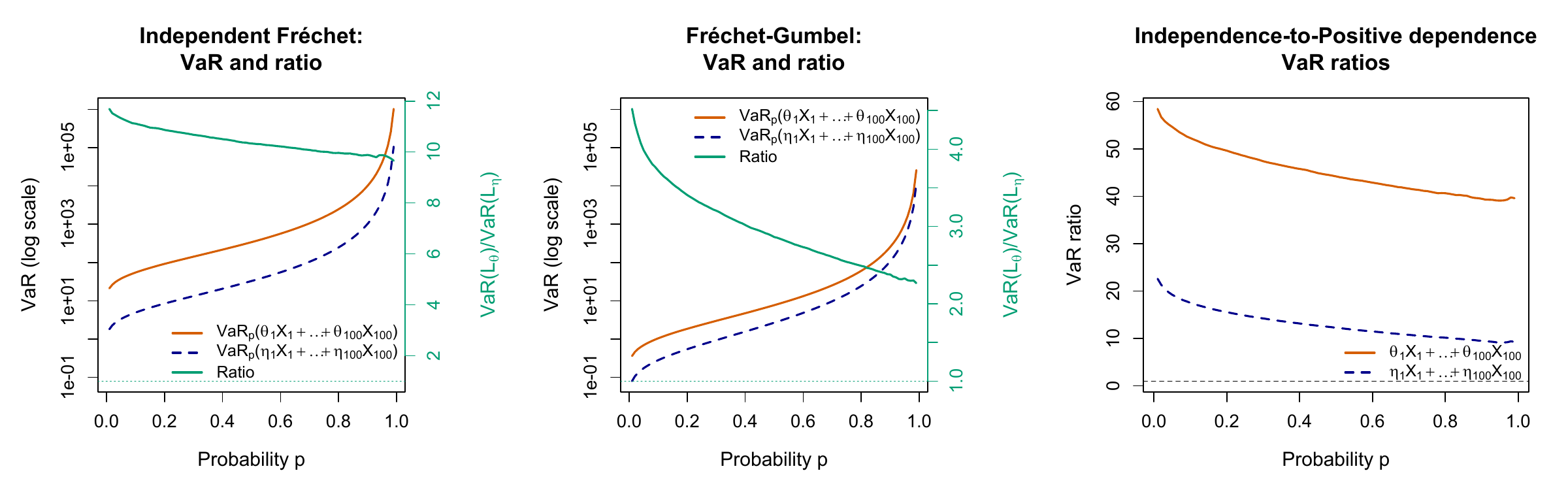}
\captionsetup{font=small}
\caption{Simulated $\VaR_p$ for $\theta_1X_1+\dots+\theta_{100}X_{100}$ and $\eta_1X_1+\dots+\eta_{100}X_{100}$ for independent $\mathrm{\text{Fr\'echet}}(1/2)$ (left), Fr\'echet-Gumbel($1/2$) model (middle), and independence-to-positive dependence VaR ratio (right) for $p\in(0,1)$.}
\label{fig:frechet_gumbel}
\end{figure}

Figure \ref{fig:frechet_gumbel} shows the diversification penalty for $n=100$ with Fr\'echet$(1/2)$ marginals under independence and Fr\'echet-Gumbel$(1/2)$ dependence by simulations. To simulate a Positive Stable$(1/2)$ random variable $R$, draw $U\sim\mathrm{Unif}(0,1)$ and $E\sim\mathrm{Exp}(1)$ independently and use $R \laweq 1/(4E\cos^2(\pi U/2))$ \citep{devroye2014simulation}. At every probability $p\in(0,1)$, the equally weighted portfolio with $(\theta_1,\dots,\theta_{100})=(0.01, \dots,0.01)$ has a higher estimated VaR than the less diversified portfolio with $(\eta_1,\dots,\eta_{100})=(0.95,0.05/99,\dots,0.05/99)$ in both models. For each portfolio, the empirical VaR is higher under independence than under Fr\'echet-Gumbel dependence at every plotted probability level $p$.

Table \ref{tab:copula-examples} presents further examples of $(X_1,\dots,X_n)$ with completely monotone Archimedean copulas that satisfy the diversification penalty; see \cite{hofert2012stochastic} and \cite{semenikhine2018multiplicative} for the corresponding generators. The copula families in Table \ref{tab:copula-examples} cover a range of positive dependence strengths. All of them except the Ali--Mikhail--Haq family permit dependence ranging from independence to arbitrarily strong positive dependence. In contrast, the Ali--Mikhail--Haq family has a pairwise Kendall's tau \footnote{See p.~158 of \cite{N06} for the definition of Kendall's tau, and p.~163 for the Archimedean formulas.} of $0<\tau<1/3$ over the stated parameter range, showing limited positive dependence.



\begin{table}[htbp]
\centering
\renewcommand{\arraystretch}{1.5}
\scriptsize
\caption{Examples of marginal distributions and  Archimedean copulas of \eqref{eq:archimedean-representation}}
\label{tab:copula-examples}
\resizebox{\textwidth}{!}{%
\begin{tabular}{ccccc}
\toprule
\textbf{Copula} & \textbf{Parameter} & \textbf{Generator $\psi(t)$} & \textbf{Background risk $R$} & \textbf{Marginal distribution $F$ of $X_i$} \\ 
\midrule

Gumbel 
& $\theta\geq1$
& $\exp(-t^{1/\theta})$
& $\mathrm{Positive\ Stable}(1/\theta)$
& Fr\'echet: $F(x)=\exp(-x^{-1/\theta})$
\\

Clayton 
& $\theta >0$
& $(1+t)^{-1/\theta}$
& $\Ga(1/\theta,1)$
& $\BP$: $F(x)=(x/(1+x))^{1/\theta}$
\\

Joe
& $\theta\geq1$
& $1-\left(1-e^{-t}\right)^{1/\theta}$
& $\mathrm{Sibuya}(1/\theta)$
& $F(x)=1-\left(1-e^{-1/x}\right)^{1/\theta}$
\\

Frank
& $\theta>0$
& $-
\log\left(1-(1-e^{-\theta})e^{-t}\right)/\theta$
& $\mathrm{Logarithmic}(1-e^{-\theta})$
& $F(x)=-\log\left(1-(1-e^{-\theta})e^{-1/x}\right)/\theta$
\\

Ali--Mikhail--Haq
& $0\leq\theta<1$
& $(1-\theta)/(e^t-\theta)$
& $\mathrm{Geometric}(1-\theta)$
& $F(x)=(1-\theta)/(e^{1/x}-\theta)$
\\

Multiplicative Multivariate Gamma
& $0<\theta<1$
& $\Gamma(\theta,t)/\Gamma(\theta)$
& $R-1\sim\BP(1-\theta,\theta)$
& Inverse Gamma: $F(x)=\Gamma(\theta,1/x)/\Gamma(\theta)$
\\

\bottomrule
\end{tabular}
}
\end{table}

\section{Pareto risks with some Archimedean copulas}
\label{sec:Pareto}
In this section, we study the diversification penalties under LRS for general background risk models \eqref{eq:general-background risk} with Pareto marginals. This discussion can be easily generalized to other heavy-tailed distributions. For $i\in[n]$, suppose that each $X_i$ has a Pareto($\alpha$) marginal distribution, where $0<\alpha\leq1$. Denote by $F_X$ the distribution function of Pareto($\alpha$). Sklar's Theorem gives the  stochastic representation
$(X_1,\dots,X_n)=(F_X^{-1}(U_1),\dots,F_X^{-1}(U_n))$
where $(U_1,\dots,U_n)$ follows the copula of $(X_1,\dots,X_n)$.  Let $(U_1,\dots,U_n)$ follow a completely monotone Archimedean copula with generator $\psi$.
We have $$(U_1,\dots,U_n)\laweq\left(\psi\left(\frac{1}{RY_1}\right),\dots,\psi\left(\frac{1}{RY_n}\right)\right),$$ for $i\in[n]$, where $1/Y_i$ follows the exponential distribution with rate 1 \citep[Proposition 7.51 (ii)]{MFE15}. Hence, $(X_1,\dots,X_n)$ has a background risk model 
\begin{equation}
\label{eq:Pareto_mixture}
(X_1,\dots,X_n)\laweq(f(Y_1,R),\dots,f(Y_n,R)),
\end{equation}
where
$f(y,r)=F_X^{-1}(\psi(1/ry))$ for $y,r>0$.

Applying Lemma \ref{thm:conditional-dp}, we derive conditions relating $\alpha$ to the parameter $\theta$ of the completely monotone Archimedean copulas in Table \ref{tab:copula-mixture-examples}, to ensure that the diversification penalty \eqref{eq:dpl} holds for Pareto marginals. Under the stated parameter relationship, the copula families in Table \ref{tab:copula-mixture-examples} cover a range of positive dependence strengths. All of them except the Joe family and the family whose survival copula is Clayton exhibit limited positive dependence, with a pairwise Kendall's tau of $0<\tau<1/3$. In contrast, the  Joe family and the family whose survival copula is Clayton accommodate arbitrarily strong positive dependence as $\alpha$ approaches zero. Denote by $\mathcal C_n$ the set of $n$-dimensional copulas. Define $\mathcal C_{n,\alpha}=\{C\in\mathcal C_n \mid C  \text{~belongs to the class of copulas in  Table \ref{tab:copula-mixture-examples}}\}$.

\begin{table}[htbp]
\centering
\renewcommand{\arraystretch}{1}
\small
\captionsetup{font=small}
\caption{Completely monotone Archimedean copulas for the diversification penalty \eqref{eq:dpl} with Pareto($\alpha$) marginals}
\label{tab:copula-mixture-examples}

\begin{tabular}{ccc}
\toprule
\makebox[0.27\textwidth][c]{\textbf{Copula}} & \makebox[0.35\textwidth][c]{\textbf{Generator $\psi(t)$}} & \makebox[0.29\textwidth][c]{\textbf{Sufficient parameter conditions}} \\ 
\midrule

Independence
& $e^{-t}$
& N/A
\\

Gumbel 
& $\exp(-t^{1/\theta})$
& $1 \leq \theta \leq (3-\alpha)/2$
\\

Joe
& $1-\left(1-e^{-t}\right)^{1/\theta}$
& $1 \leq \theta \leq 1/\alpha$
\\

Clayton 
& $(1+t)^{-1/\theta}$
& $0<\theta \leq1$
\\

Ali--Mikhail--Haq
& $(1-\theta)/(e^t-\theta)$
& $0\leq\theta<1$
\\

Frank
&
$-\log\!\left(1 - (1 - e^{-\theta})e^{-t}\right)/\theta$
& $0<\theta \leq 1.5$
\\

Survival copula: Clayton 
& $(1+t)^{-1/\theta}$
& $\theta=1/\alpha$
\\

\bottomrule
\end{tabular}
\end{table}

\begin{theorem}
\label{thm:Pareto-mixture}
Fix a positive integer $m$.
Suppose that $(X_1,\dots, X_n)$ has Pareto$(\alpha)$ marginal risks with $\alpha\in(0,1]$, and copula  $C=\sum_{j=1}^mp_jC_j$ where $C_j\in \mathcal C_{n,\alpha}$ for all $j\in[m]$  and $(p_1,\dots,p_m)\in\Delta_m$. Then for any $(\theta_1,\dots, \theta_n)$ and $(\eta_1,\dots, \eta_n)$ in $\R_+^n$ with $(\theta_1, \dots, \theta_n)\preceq(\eta_1, \dots, \eta_n)$, 
\[
\eta_1X_1+\dots+\eta_nX_n \leq_{\st} \theta_1X_1+\dots+\theta_nX_n.
\]

\end{theorem}

See Section 1 of the Supplementary Materials for the proof. Under the parameter conditions in Table \ref{tab:copula-mixture-examples}, Pareto risks with a common tail parameter $\alpha$ satisfy the diversification penalty \eqref{eq:dpl}. We will now consider the diversification penalty \eqref{eq:dpl2} for Pareto marginals with different tail parameters.
\begin{proposition}
Fix a positive integer $m$. For $i\in[n]$,
let $X_i \sim\mathrm{Pareto}(\alpha_i)$  with $\alpha_i\in(0,1]$. Write $\alpha^*=\max_{i\in[n]}\{\alpha_i\}$. Suppose that the copula of $(X_1,\dots,X_n)$ is $C=\sum_{j=1}^mp_jC_j$ where $C_j\in \mathcal C_{n,\alpha^*}$ for all $j\in[m]$  and $(p_1,\dots,p_m)\in\Delta_m$. For all  $p\in(0,1)$ and $(w_1,\dots,w_n)\in\Delta_n$, $$\VaR_p(w_1X_1+\dots+w_nX_n)\ge \VaR_p(w_1X_1)+\dots+\VaR_p(w_nX_n).$$ 
\end{proposition}
\begin{proof}
For $i\in[n]$, suppose that $(Y_1,\dots,Y_n)$ has a common Pareto($\alpha^*$) marginal with a copula $C\in\mathcal C_{n,\alpha^*}$. By Theorem \ref{thm:Pareto-mixture}, $Y_1\leq_{\mathrm{st}}\sum_{i=1}^n w_iY_i$ for any $(w_1,\dots,w_n)\in\Delta_n$. This stochastic relation is equivalent to the weighted universal VaR superadditivity of $(Y_1,\dots,Y_n)$; see Property \eqref{eq:dpl2}. For $0<\alpha_i\leq\alpha^*$ and $i\in[n]$, define $X_i=(1+Y_i)^{\alpha^*/\alpha_i}-1$. Hence, $X_i \sim \mathrm{Pareto(\alpha_i)}$. The strictly increasing transformation from $Y_i$ to $X_i$ preserves the copula $C$ \citep[Proposition 7.7]{MFE15}. The transformation is also convex, thus Theorem 1 (i) of \cite{CLW26} and the positive homogeneity of VaR imply the desired result. 
\end{proof}

\section{Conditional mean risk sharing}
\label{sec:CMRS}
Consider the same risk sharing pool introduced at the start of Section \ref{sec:LRS}: agent $i\in[n]$ faces a risk $a_iX_i$, where $a_i>0$ and $(X_1,\dots,X_n)$ follows the MBR model \eqref{eq:MBR}, and  the total loss of the pool is  $S=\sum_{i=1}^n a_iX_i$. Diversification penalties under LRS rule \eqref{eq:dpl} and \eqref{eq:dpl2} have been established in Theorem \ref{thm:common-multiplier} when the background risk is IC. We will show that a diversification penalty under the conditional mean risk sharing rule also holds with an IC background risk.  
 
The \emph{conditional mean risk sharing (CMRS)} \citep{denuit2012convex}  defines the loss position of agent $i\in[n]$ after risk sharing by $m_i(S)=\mathbb{E}[a_iX_i\mid S]$. The full allocation $\sum_{i=1}^n m_i(S)=S$ holds almost surely. 
In many probability textbooks, the conditional expectation is often defined for integrable random variables (e.g., p.~178 of \cite{D19}). To accommodate non-negative random variables $X_1,\dots,X_n$ with infinite means, we define the CMRS as
\[
m_i(S)=\mathbb{E}[a_iX_i\mid S]:=\lim_{k\to\infty}\mathbb{E}[a_iX_i\wedge k\mid S], \qquad i\in[n].
\]
Since $X_1,\dots,X_n$ are non-negative and each $a_iX_i\wedge k$ is integrable, the CMRS is well defined by the monotone convergence theorem. 
Moreover, $0\le m_i(S)\le S<\infty$ holds almost surely as $S<\infty$ holds almost surely.

For finite-mean losses with an arbitrary dependence structure, CMRS is beneficial for every agent $i\in[n]$ under convex order, namely $m_i(S)\le_{\mathrm{cx}}a_iX_i$ \citep[Proposition 4.2]{denuit2012convex}. It means that CMRS preserves each agent's expected loss while reducing the variability of their individual loss.  Moreover, CMRS exhibits a linear allocation form in several important models. For example, exchangeability of $(a_1X_1,\ldots,a_nX_n)$ yields $m_i(S)=S/n$ for every agent $i\in[n]$. Linear allocations also arise in the Gamma model with a common rate discussed in Example \ref{ex:linear_CMRS}, and the stable model examined in Theorem \ref{thm: CMRS_stable}. Although these allocations are linear, CMRS needs not preserve each agent's risk exposures before and after risk sharing, as required by LRS.

As noted in Remark 2.4 of \cite{denuit2025comonotonicity}, CMRS can be detrimental for risks with infinite mean. Assume that $a_1=\dots=a_n$ and $X_1,\dots,X_n\sim\mathrm{Pareto}(\alpha)$ are iid with $\alpha\in (0,1]$. Then $m_i(S)=S/n$ coincides with LRS. By Theorem 1 of \cite{chen2025unexpected}, 
\begin{equation}
\label{eq:CMRSdp}
a_iX_i\le_{\rm st}m_i(S)\quad\text{for all}\quad i\in[n].
\end{equation}
We say that a loss vector $(X_1,\dots,X_n)$ satisfies the \emph{diversification penalty under CMRS} if  \eqref{eq:CMRSdp}  holds. The following theorem establishes the diversification penalty beyond iid risks.

\begin{theorem}
\label{thm:inverted-concave-common-factor}
For $i\in[n]$, suppose that $X_i=RY_i$ where $Y_1,\dots,Y_n$ are positive random variables with a joint density, and $R\in\mathrm{IC}$ is independent of $(Y_1,\dots,Y_n)$. For  $(a_1,\dots,a_n)\in\R_+^n$, let $S=\sum_{i=1}^na_iX_i$. Then $a_iX_i\le_{\st}m_i(S)$ for every $i\in[n]$.
\end{theorem}

\begin{proof}

According to Theorem 1.A.1 of \cite{SS07}, for every $i\in[n]$, $a_iX_i\le_{\mathrm{st}}m_i(S)$ if and only if there exist $a_iP\laweq a_iX_i$ and $Q\laweq m_i(S)$ such that $\p(a_iP\le Q)=1$. Let $V\sim\operatorname{Unif}(0,1)$ and define $X^*_i=F_{X_i}^{-1}(V)$ and $S^*=F_S^{-1}(V)$. By the quantile transformation \citep[Proposition 7.2]{MFE15}, $X^*_i\laweq X_i$ and $S^*\laweq S$. Hence, it suffices to show that $a_iX_i^*\le m_i(S^*)$ holds almost surely. We will show $a_iF_{X_i}^{-1}(p)\le m_i(F_S^{-1}(p))$ for all $p\in(0,1)$.

Write $\bm{w}=(w_1,\dots,w_n)$. Let $S_{\bm{w}}=\sum_{i=1}^nw_iX_i$ and $Q_p(\bm{w})$ be the $p$-quantile of $S_{\bm{w}}$. Denote by $\mathbf{e}_i$  the vector whose $i$-th coordinate is one and whose other coordinates are zero. By Theorem \ref{thm:common-multiplier}, $Q_p$ is concave. Then, define $\boldsymbol{a}=(a_1,\dots,a_n)$, we have
\[
F_{X_i}^{-1}(p)=Q_p(\mathbf{e}_i)\le Q_p(\boldsymbol{a})+\sum_{j=1}^n \left((\mathbf{e}_i)_j-a_j \right)\frac{\partial Q_p(\boldsymbol{a})}{\partial a_j}=Q_p(\boldsymbol{a})+\frac{\partial Q_p(\boldsymbol{a})}{\partial a_i}-\sum_{j=1}^n a_j\frac{\partial Q_p(\boldsymbol{a})}{\partial a_j}=\frac{\partial Q_p(\boldsymbol{a})}{\partial a_i}.
\]
The last equality is due to Euler's identity \citep[Equation 8.61]{MFE15} 
\[
Q_p(\boldsymbol{a})=\sum_{j=1}^n a_j\frac{\partial Q_p(\boldsymbol{a})}{\partial a_j}.
\]
Since $Y_1,\dots,Y_n$ have a joint density and $R>0$ almost surely, $X_1,\dots,X_n$ also have a joint density. Then, for a fixed $p\in(0,1)$, $Q_p$ is differentiable at $\boldsymbol{a}$  \citep{gourieroux2000sensitivity}.  Equation 6 of \cite{tasche2001conditional} yields
\[
\frac{\partial Q_p(\bm{a})}{\partial a_i}=\frac{m_i({F_{S}^{-1}(p)})}{a_i}.
\]
Therefore, for all $p\in(0,1)$, 
$$
m_i({F_{S}^{-1}(p)})=a_i \frac{\partial Q_p(\bm{a})}{\partial a_i} \geq a_i Q_p(\mathbf{e}_i)=a_iF_{X_i}^{-1}(p).$$
The proof is complete. 
\end{proof}

\begin{corollary}
\label{cor:frechet-gumbel}
 For  $(a_1,\dots,a_n)\in\R_+^n$, let $S=\sum_{i=1}^na_iX_i$. Then $a_iX_i\le_{\st}m_i(S)$  holds for every $i\in[n]$ if one of the following conditions holds.
\begin{enumerate}[(i)]
\item $(X_1,\dots,X_n)$ follows the Pareto-Clayton($\alpha$) model \eqref{eq:PC} with $0<\alpha \leq1$;
\item $(X_1,\dots,X_n)$ follows the Fr\'echet-Gumbel($\delta$) model \eqref{eq:FG} with $0<\delta \leq 1/2$.
\end{enumerate}
\end{corollary}

\begin{proof}
It is clear that the statement holds under (i). We show it under  (ii).
Since $Y_1,\ldots,Y_n$ are iid Fr\'echet$(1)$ random variables, $Y_1,\dots,Y_n$ have a joint density.
Next, we show $R\in\mathrm{IC}$. For $u>0$, define $g(u)=\mathbb{P}(R>1/u)$. The error function is defined by $\operatorname{erf}(x)=(2/\sqrt{\pi})\int_0^x e^{-t^2}\,dt$ for all $x\in \R$. For $x>0$, it is increasing and concave in $x$. When $\delta=1/2$, $R$ follows the L\'evy distribution with location $0$ and scale $1/2$, and has the cumulative distribution function of $1-\operatorname{erf}(1/(2\sqrt{u}))$ for $u>0$. Hence, $g(u)=\operatorname{erf}(\sqrt{u}/2)$ is concave in $u$. For $0<\delta<1/2$, let $W$ and $V$ be independent Positive Stable random variables with parameters $1/2$ and $2\delta$, respectively. Then, we have
\[
\mathbb{E}[e^{-tW^{1/(2\delta)}V}]=\mathbb{E}\left[\mathbb{E}\left[e^{-tW^{1/(2\delta)}V}\mid W\right]\right]=\mathbb{E}\left[\exp\left\{-\left(tW^{1/(2\delta)}\right)^{2\delta}\right\}\right]=\mathbb{E}[e^{-t^{2\delta}W}]=e^{-(t^{2\delta})^{1/2}}=e^{-t^\delta}.
\]
Hence, the uniqueness of the Laplace transform implies $R\laweq W^{1/(2\delta)}V$. Based on the independence of $W$ and $V$, for $u>0$, we obtain
\[
g(u)=\mathbb{P}(R>1/u)=\mathbb{E}\left[\mathbb{P}\left(W^{1/(2\delta)}>\frac{1}{uV}\,\middle|\,V\right)\right]=\mathbb{E}\left[\mathbb{P}\left(W>(uV)^{-2\delta}\,\middle|\,V\right)\right]=\mathbb{E}\left[\operatorname{erf}\left(\frac{(uV)^\delta}{2}\right)\right].
\]
Here, $g(u)$ is concave as $\operatorname{erf}(u)$ is increasing and concave on $u \in (0,\infty)$, and the transformation $u\mapsto(uv)^\delta/2$ is increasing and concave for $0<\delta<1/2$. Hence, $R\in\mathrm{IC}$. By Theorem \ref{thm:inverted-concave-common-factor}, the desired CMRS stochastic order result is obtained.
\end{proof}



We present below another example for which the diversification penalty \eqref{eq:CMRSdp} holds.

\begin{example}[Common rate idiosyncratic Gamma risks]
\label{ex:linear_CMRS} 
For $i\in[n]$, let $a_i=1$ and suppose that $X_i=RY_i$ where $Y_i\sim\operatorname{Gamma}(k_i,\lambda)$ with $k_i>0$ and $\lambda>0$, $Y_1,\dots,Y_n$ are independent, and $R\in\mathrm{IC}$ is independent of $(Y_1,\dots,Y_n)$. Thus,  the idiosyncratic risks share the common rate $\lambda$.

For $i\in[n]$, define $P_i=Y_i/T$ where $T=\sum_{i=1}^n Y_i$, and $K=\sum_{i=1}^n k_i$. Hence, we have $S=\sum_{i=1}^n X_i=RT$. The Gamma-Dirichlet decomposition \citep[proof of Theorem 2.1]{ng2011dirichlet} implies that $(P_1,\dots,P_n)$ is independent of $T$. Thus,  $\mathbb{E}[P_i]=\E[Y_i/T]=\E[Y_i]/\E[T]=k_i/K$. Moreover, $R$ is independent of $(Y_1,\dots,Y_n)$, meaning that $R$ is independent of $(P_1,\dots,P_n,T)$, since $(P_1,\dots,P_n,T)$ is a function of $(Y_1,\dots,Y_n)$. Therefore, for $i\in[n]$, $P_i$ is independent of $(R,T)$, and is also independent of $S=RT$. Finally, we have
\[
m_i(S)=\mathbb{E}[X_i\mid S]=\mathbb{E}[SP_i\mid S]=S\,\mathbb{E}[P_i\mid S]=S\,\mathbb{E}[P_i]=\frac{k_i}{K}S,\qquad i\in[n].
\]
Since $Y_1, \dots, Y_n$ are mutually independent Gamma random variables, $Y_1, \dots, Y_n$ have a joint density. Since $R\in\mathrm{IC}$, Theorem \ref{thm:inverted-concave-common-factor} gives $a_iX_i\le_{\mathrm{st}}m_i(S)$.
\end{example}

\section{Conditional mean risk sharing for stable risks}
\label{sec:CMRS_Stable}
In this section, we derive the CMRS expression for independent stable idiosyncratic risks, and discuss peakedness order and stochastic order relations, under different parameter conditions. We first provide the definition of peakedness order.

\begin{definition}{\citep[][p.~178]{SS07}}
\label{def:peakedness}
For random variables $X$ and $Y$ that are both symmetric about zero, peakedness order about zero is denoted by $X\le_{\mathrm{peak}}Y$ when $|X|\le_{\mathrm{st}}|Y|$, equivalently $\p(|X|\le t)\ge\p(|Y|\le t)$ for all $t\ge0$.
\end{definition}
It means that $X$ is more concentrated about zero than $Y$, with no greater probability of exceeding any given threshold in absolute value. For non-negative losses, it coincides with usual stochastic order. \cite{ibragimov2005new} shows that under LRS, for iid $\alpha$-stable risks $X_1,\dots,X_n$ with location parameter zero, more equal portfolio weights increase peakedness when $1<\alpha\le2$, but decrease it when $0<\alpha<1$. Equivalently, let $(\theta_1,\dots,\theta_n)$ and $(\eta_1,\dots,\eta_n) \in \R_+^n$ with $(\theta_1,\dots,\theta_n)\preceq(\eta_1,\dots,\eta_n)$. Then $\sum_{i=1}^n \theta_iX_i \le_{\rm peak} \sum_{i=1}^n \eta_iX_i$ when $1<\alpha\le2$, and $\sum_{i=1}^n \eta_iX_i \le_{\rm peak} \sum_{i=1}^n \theta_iX_i$ when $0<\alpha<1$. This celebrated result has found numerous applications in economics \citep{ibragimov2015heavy}. We will show that a similar peakedness order relation also holds for stable risks under CMRS.


We use the $S_1$ parameterization of stable distributions, following \cite{samorodnitsky1994stable}. Denote by $S(\alpha,\beta,\sigma,\mu)$ the stable distribution, with stability parameter $\alpha \in (0,2]$, skewness parameter $\beta\in[-1,1]$, scale parameter $\sigma>0$, and location parameter $\mu\in\mathbb{R}$.  The characteristic function of $Y \sim S(\alpha,\beta,\sigma,\mu)$ is given by, for $y \in \mathbb{R}$,
\begin{equation}
\varphi_Y(y)=\E[e^{\mathrm{i} yY}]=
\begin{cases}
\displaystyle
\exp\!\left\{
\mathrm{i}\mu y-\sigma^\alpha|y|^\alpha
\left(
1-\mathrm{i}\beta \operatorname{sgn}(y)\tan\!\left(\frac{\pi\alpha}{2}\right)
\right)
\right\},
& \alpha\ne1,\\[1.2ex]
\displaystyle
\exp\!\left\{
\mathrm{i}\mu y-\sigma|y|
\left(
1+\mathrm{i}\frac{2\beta}{\pi}\operatorname{sgn}(y)\log|y|
\right)
\right\},
& \alpha=1.
\end{cases}
\label{eq:stable-definition}
\end{equation}
Here, $\mathrm{i}=\sqrt{-1}$ and $\operatorname{sgn}(\cdot)$ is the sign function. Most stable random variables do  not have explicit distribution functions, except the normal distribution ($\alpha=2$), the Cauchy distribution ($\alpha=1$ and $\beta=0$), and the L\'evy distribution ($\alpha=1/2$ and $\beta=1$). Note that if  $\alpha > 1$,  the mean of the stable distribution is finite, and if $\alpha \le 1$, the mean of the stable distribution can be $+\infty$, $-\infty$, or not defined. Note that non-degenerate stable distributions have unimodal and bounded densities; the boundedness is implied by Proposition 28.1 and Example 28.2 of \cite{S99}. If $\alpha\ne1$ or $\beta=0$, a useful property for stable distributions is that $\sum_{i=1}^n a_iY_i\sim S(\alpha,\beta,(\sum_{i=1}^n(a_i\sigma_i)^\alpha)^{1/\alpha},0)$ where $Y_1,\dots,Y_n \sim S(\alpha,\beta,\sigma_i,0)$ are independent and $(a_1,\dots,a_n)\in\R_+^n$ \cite[Property 1.2.1]{samorodnitsky1994stable}.

Under suitable conditions on the joint stable distribution, conditional absolute moments of stable distributions can be finite even when the unconditional moments are infinite \citep[Theorem 5.1.3]{samorodnitsky1994stable}.
Theorem 3.1(ii) of \cite{samorodnitsky1991conditional} established conditions under which a jointly symmetric $\alpha$-stable pair $(U,V)$ satisfies $\E[V\mid U=u]=\kappa u$ for almost every $u$. Similar to this structure, Theorem \ref{thm: CMRS_stable} derives explicit CMRS allocations for weighted independent stable distributions in the MBR models \eqref{eq:MBR} under both finite-mean and infinite-mean settings. We first show that for independent stable risks, the CMRS is always finite. 
For $i\in[n]$, let $Y_i\sim S(\alpha,\beta,\sigma,0)$ be independent, and $S=\sum_{i=1}^n Y_i$. Define $U_i=\sum_{j\ne i}Y_j$. Then $U_i$ is independent of $Y_i$ and has a bounded unimodal density $f_{U_i}$. Hence, for every $s\in\R$, we have
\[
\mathbb{E}[|Y_i|\mid S=s]=   \frac{\displaystyle\int_{\mathbb{R}}|y|f_{Y_i}(y)f_{U_i}(s-y)\,dy}{\displaystyle\int_{\mathbb{R}}f_{Y_i}(y)f_{U_i}(s-y)\,dy}.
\]
By unimodality, $f_{U_i}$ is at least $f_{U_i}(x)$ on the interval between the mode $m$ and any point $x\in\mathbb{R}$. The area under the density on that interval is at most one, i.e., $|x-m|f_{U_i}(x)\leq 1$. For fixed $s$ and take $x=s-y$, we have $|y|f_{U_i}(s-y)\leq (|s-y-m|+|s-m|)f_{U_i}(s-y)\leq 1+|s-m|\sup_{x\in\mathbb{R}} f_{U_i}(x)$. The last bound is finite and independent of $y$. Thus $\mathbb{E}[|Y_i|\mid S=s]$ is always finite even when $Y_i$ is not integrable.

\begin{theorem}
\label{thm: CMRS_stable}
For $i\in[n]$, suppose that $X_i=RY_i$ where $Y_1,\dots,Y_n \sim S(\alpha,\beta,\sigma_i,0)$ are independent with  $\sigma_i>0$, and $R>0$ is independent of $(Y_1,\dots,Y_n)$. For $(a_1,\dots,a_n)\in\R_+^n$, let $S=\sum_{i=1}^na_iX_i$. Then for every $i\in[n]$, if either $\alpha\in(0,1)\cup(1,2]$ and $\beta\in[-1,1]$ or $\alpha=1$ and $\beta=0$ holds,
$$m_i(S)=\frac{(a_i\sigma_i)^\alpha}{\sum_{i=1}^n (a_i\sigma_i)^\alpha} S.$$ 
\end{theorem}

\begin{proof}
We first consider the case when $R=1$. The case when $R$ is non-degenerate is direct by independence.
For $i\in[n]$, define $B_i=a_iY_i$, $T=\sum_{i=1}^n B_i$, $c_i=(a_i\sigma_i)^\alpha$ and $C=\sum_{i=1}^n c_i$. Thus, $a_iX_i=RB_i$ and $S=RT$.  The characteristic function of $B_i$ is, for $u\in\mathbb{R}\setminus\{0\}$,
\[
\varphi_{B_i}(u)=\E[e^{\mathrm{i}uB_i}]=\E[e^{\mathrm{i}(a_iu)Y_i}]=\exp\!\left\{-c_i|u|^\alpha \bigl[1-\mathrm{i}\tau\operatorname{sgn}(u)\bigr]\right\},
\]
where  $\tau=\beta\tan(\pi\alpha/2)$ when $\alpha\ne1$, and $\tau=0$ when $\alpha=1$ and $\beta=0$.
By independence of $(Y_1, \dots, Y_n)$, we have
\[
\varphi_T(u)=\prod_{i=1}^n\varphi_{B_i}(u)=\exp\!\left\{-C|u|^\alpha\bigl[1-\mathrm{i}\tau\operatorname{sgn}(u)\bigr]\right\}.
\]

Now, for $i\in[n]$, define $U_i=\sum_{j\ne i}B_j$. Then $T=B_i+U_i$, and $B_i$ and $U_i$ are independent. Additionally, the characteristic function of $U_i$ is, for $u\in\mathbb{R}\setminus\{0\}$, 
\[
\varphi_{U_i}(u)=\exp\!\left\{-(C-c_i)|u|^\alpha\bigl[1-\mathrm{i}\tau\operatorname{sgn}(u)\bigr]\right\}.
\]
For $t\in\mathbb{R}$, 
$$\E[B_i\mid T=t]=h_i(t)/f_T(t) \quad\text{where} \quad h_i(t)=\int_{\mathbb{R}}x f_{B_i}(x)f_{U_i}(t-x)\,dx.$$

The characteristic functions are integrable. We can therefore use the Fourier inversion to recover the densities from the characteristic functions. Thus, $f_{B_i}(x)=
\int_{\mathbb{R}}
e^{-\mathrm{i} ux}\varphi_{B_i}(u)/(2\pi)\,du$. Integration by parts gives
\[
\int_{\mathbb{R}}e^{-\mathrm{i} ux}\varphi_{B_i}'(u)\,du=\left[e^{-\mathrm{i} ux}\varphi_{B_i}(u)\right]_{-\infty}^{\infty}+\mathrm{i} x \int_{\mathbb{R}}e^{-\mathrm{i} ux}\varphi_{B_i}(u)\,du=2\pi\mathrm{i} x f_{B_i}(x),
\]
since the boundary terms are zero. Therefore,
\begin{equation}
x f_{B_i}(x)=\frac{1}{2\pi\mathrm{i}}\int_{\mathbb{R}}e^{-\mathrm{i} ux}\varphi_{B_i}'(u)\,du.
\label{eq:weighted-density-inversion}
\end{equation}
The same identity applies to $t f_T(t)$, and we have $t f_{T}(t)=\int_{\mathbb{R}}e^{-\mathrm{i} ut}\varphi_{T}'(u)/(2\pi\mathrm{i})\,du$. Substituting \eqref{eq:weighted-density-inversion} into $h_i(t)$, we obtain
\begin{align*}
h_i(t)
&=
\frac{1}{2\pi\mathrm{i}}
\int_{\mathbb{R}}
\left[
\int_{\mathbb{R}}
e^{-\mathrm{i} ux}\varphi_{B_i}'(u)\,du
\right]
f_{U_i}(t-x)\,dx =
\frac{1}{2\pi\mathrm{i}}
\int_{\mathbb{R}}
\varphi_{B_i}'(u)
\left[
\int_{\mathbb{R}}
e^{-\mathrm{i} ux}f_{U_i}(t-x)\,dx
\right]du\\
&=\frac{1}{2\pi\mathrm{i}}
\int_{\mathbb{R}}
\varphi_{B_i}'(u)
\left[e^{-\mathrm{i} ut}\int_{\mathbb{R}}
e^{\mathrm{i} uz}f_{U_i}(z)\,dz
\right]du \qquad (\text{Let }z=t-x)\\
&=\frac{1}{2\pi\mathrm{i}}
\int_{\mathbb{R}}
e^{-\mathrm{i} ut}
\varphi_{B_i}'(u)\varphi_{U_i}(u)\,du.
\end{align*}

For $u\ne0$, differentiating $\varphi_{B_i}(u)$ and $\varphi_T(u)$ gives
\[
\varphi_{B_i}'(u)=-\alpha c_i|u|^{\alpha-1}\left(\operatorname{sgn}(u)-\mathrm{i}\tau\right)\exp\!\left\{-c_i|u|^\alpha \left[1-\mathrm{i}\tau\operatorname{sgn}(u)\right]
\right\},
\]
\[
\varphi_{T}'(u)=-\alpha C|u|^{\alpha-1}\left(\operatorname{sgn}(u)-\mathrm{i}\tau\right)\exp\!\left\{-C|u|^\alpha \left[1-\mathrm{i}\tau\operatorname{sgn}(u)\right]
\right\}.
\]
Therefore, we have
\[
\varphi_{B_i}'(u)\varphi_{U_i}(u)=-\alpha c_i|u|^{\alpha-1}\left(\operatorname{sgn}(u)-\mathrm{i}\tau\right)\exp\!\left\{-C|u|^\alpha \left[1-\mathrm{i}\tau\operatorname{sgn}(u)\right]\right\}=\frac{c_i}{C}\varphi_T'(u).
\]
Applying \eqref{eq:weighted-density-inversion} to $T$, we have
\[
h_i(t)=\frac{c_i}{C}\frac{1}{2\pi\mathrm{i}}\int_{\mathbb{R}}e^{-\mathrm{i}ut}\varphi_T'(u)\,du=\frac{c_i}{C}\,t f_T(t).
\]
Hence, $\E[B_i\mid T=t]=h_i(t)/f_T(t)=(c_i/C)t$. For $r>0$, observing $S=s$ and $R=r$ determines $T=s/r$. Independence of $R$ from $(B_i,T)$ implies
\[
\E[RB_i\mid S=s,R=r]=r\,\E[B_i\mid T=s/r,R=r]=r\,\E[B_i\mid T=s/r]=r\,\frac{c_i}{C}\frac{s}{r}=\frac{c_i}{C}s.
\]
Finally, we obtain
\[
m_i(S)=\E[a_iX_i\mid S]=\E[RB_i\mid S]=\E\!\left[\E[RB_i\mid S,R]\mid S\right]=\E\!\left[\frac{c_i}{C}S\mid S\right]=\frac{c_i}{C}S=\frac{(a_i\sigma_i)^\alpha}{\sum_{i=1}^n(a_i\sigma_i)^\alpha}\,S.
\]
This completes the proof.
\end{proof}

When $\alpha=1$ and $\beta\ne0$, the logarithmic term in the $S_1$ characteristic function \eqref{eq:stable-definition} introduces a location shift for $a_iY_i$ based on the weight $a_i$ for every $i\in[n]$. Hence, an additional correction term will appear in the CMRS allocation. Although the CMRS allocation in Theorem \ref{thm: CMRS_stable} is linear in $S$, the coefficient vectors, representing an agent's exposures, generally have different sums before and after risk sharing. Therefore, the standard equal-sum majorization results cannot be applied directly. Theorem \ref{thm: peakedness} compares each agent's initial risk $a_iX_i$ with their CMRS allocation $m_i(S)$ under peakedness order.

\begin{theorem}
\label{thm: peakedness}
For $i\in[n]$, suppose that $X_i=RY_i$ where $Y_1,\dots,Y_n\sim S(\alpha,\beta,\sigma_i,0)$ are independent with $\sigma_i>0$, and $R>0$ is independent of $(Y_1, \dots, Y_n)$. For $(a_1,\dots,a_n)\in\R_+^n$, let $S=\sum_{i=1}^na_iX_i$. Then the following peakedness order relations hold for every $i\in[n]$:
\begin{enumerate}[(i)]
    \item If $0<\alpha<1$ and $ -1 \leq\beta\leq1$, then $a_iX_i\le_{\mathrm{peak}}m_i(S)$;
    \item If $\alpha=1$ and $\beta=0$, then $a_iX_i\laweq m_i(S)$;
    \item If $1<\alpha\le2$ and  $ -1 \leq\beta\leq1$, then $m_i(S)\le_{\mathrm{peak}}a_iX_i$.
\end{enumerate}
\end{theorem}

\begin{proof}
It suffices to consider the case when $R=1$.
Let $Z\sim S(\alpha,\beta,1,0)$ be a unit-scale stable risk. Define $c_i=(a_i\sigma_i)^\alpha$ and $C=\sum_{i=1}^n c_i$. Then, for $i\in[n]$, $B_i=a_iY_i\laweq c_i^{1/\alpha}Z$ and $T=\sum_{i=1}^nB_i\laweq C^{1/\alpha}Z$. Thus $|B_i|\laweq (c_i/C)^{1/\alpha}|T|$ and $m_i(S)\laweq (c_i/C)|T|$. Since $|T|$ is non-negative and $0<c_i/C\le 1$, it is clear that  $|B_i|\le_{\rm st}|m_i(S)|$ if $\alpha<1$, $|B_i|\ge_{\rm st}|m_i(S)|$ if $\alpha>1$, and $|B_i|\laweq|m_i(S)|$ if $\alpha=1$ and $\beta=0$.
\end{proof}

The following result follows directly since for $i\in[n]$ and $0<\alpha<1$, each $X_i$ is positive when $\beta=1$, and negative when $\beta=-1$.
\begin{corollary}
\label{cor: CMRS_stochastic}
For $i\in[n]$, suppose that $X_i=RY_i$ where $Y_1,\dots,Y_n\sim S(\alpha,\beta,\sigma_i,0)$ are independent with $\sigma_i>0$, and $R>0$ is independent of $(Y_1, \dots, Y_n)$.  For $(a_1,\dots,a_n)\in\R_+^n$, let $S=\sum_{i=1}^na_iX_i$. Then the following stochastic order relations hold for every $i\in[n]$:
\begin{enumerate}[(i)]
    \item If $0<\alpha<1$ and $\beta=1$, then $a_iX_i\le_{\mathrm{st}}m_i(S)$;
    \item If $\alpha=1$ and $\beta=0$, then $a_iX_i\laweq m_i(S)$;
    \item If $0<\alpha<1$ and $\beta=-1$, then $m_i(S)\le_{\mathrm{st}}a_iX_i$.
\end{enumerate}
\end{corollary}

\section{A simulation study}
\label{sec:numerical}

In this section, we compare the VaRs of LRS and CMRS allocations under a Pareto-Clayton model. Consider a pool of $n=3$ agents, and let the total exposure of the pool be 1. The three agents have initial exposures of $a_1=0.2$,
$a_2=0.3$, and $a_3=0.5$. Agent $i$ has an initial loss of $a_iX_i$, and the total loss of the pool is $S=\sum_{i=1}^{3}a_iX_i=0.2X_1+0.3X_2+0.5X_3$. We simulate $R=10^8$ independent realizations of $(X_1,X_2,X_3)$ from the Pareto-Clayton$(0.9)$ model \eqref{eq:PC}. 

We consider two LRS arrangements. The first is the equally weighted portfolio $A_i^{\mathrm{eq}}=(a_i/3)\sum_{j=1}^{3}X_j$ for a fixed agent $i\in[n]$; we do not require all agents to hold their respective benchmarks $A_i^{\mathrm{eq}}$ simultaneously.   We mainly use $A_i^{\mathrm{eq}}$ as a benchmark, as according to Theorem \ref{thm:common-multiplier}, it is the largest in usual stochastic order among all LRS arrangements for agent $i$. Under the second LRS rule, each agent retains $70\%$ of their initial loss and contributes the remaining $30\%$ to the pool. The pooled loss $0.3S$ is then shared in proportion to the agents' initial exposures, giving the final allocation of $A_i^{\mathrm{gen}}=0.7a_iX_i+0.3a_iS$.  Under $A_i^{\mathrm{gen}}$, the losses allocated to agents 1, 2, and 3 are given by 
{\small
\[
A_1^{\mathrm{gen}}=0.152X_1+0.018X_2+0.03X_3,\quad A_2^{\mathrm{gen}}=0.018X_1+0.237X_2+0.045X_3,\quad A_3^{\mathrm{gen}}=0.03X_1+0.045X_2+0.425X_3.
\]
}
For each $p\in[0.1,0.9]$ with a step size of $0.01$, we estimate the $\VaR_p$ of the initial loss $a_iX_i$, the benchmark $A_i^{\mathrm{eq}}$, and the $A_i^{\mathrm{gen}}$ allocation using their respective empirical $p$-quantiles across the simulations.



Under CMRS, agent $i$'s allocation is $m_i(S)=\E[a_iX_i\mid S]$. Following Example 4.14 of \cite{frees2025constructing}, we estimate $m_i$ from simulated losses. Denote by $(X^r_1,X^r_2,X^r_3)$ the $r$-th simulation of $(X_1,X_2,X_3)$ for all $r\in[R]$. Let $S^r=\sum_{i=1}^na_iX^r_i$ be the total loss in the $r$-th simulated path. We estimate $m_i(S^r)$ by local averaging over a symmetric window with one-sided bandwidth $h=0.02$: for $i\in[n]$ and $r\in[R]$,
\[
\hat m_i(S^r)=\frac{\sum_{\ell=1}^R a_iX_i^\ell \id_{\{S^r-h\le S^\ell\le S^r+h\}}}{\sum_{\ell=1}^R \id_{\{S^r-h\le S^\ell\le S^r+h\}}}.
\]
A sample of estimated CMRS allocations $m_i$ is thus $(\hat m_i(S^1),\dots,\hat m_i(S^R))$. For each $p\in[0.1,0.9]$ with a step size of $0.01$, we take the $p$-quantile of the sample as the estimate of $\VaR_p(m_i(S))$.

\begin{figure}[htbp]
\centering
\includegraphics[width=1\textwidth]{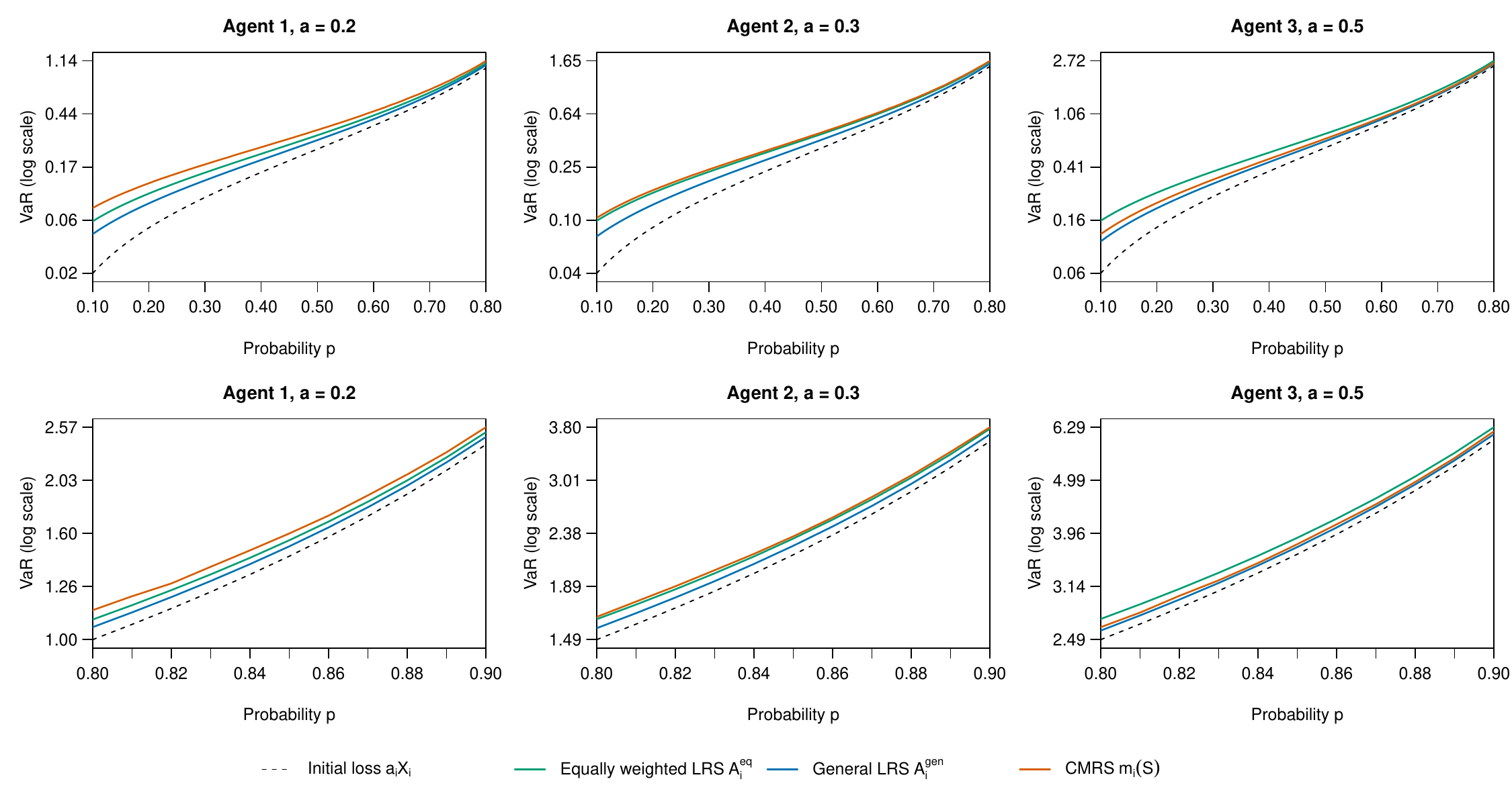}
\captionsetup{font=small}
\caption{Simulated VaR$_p$ of initial loss $a_iX_i$, equally weighted LRS allocation $A_i^{\rm eq}$,  LRS allocation $A_i^{\rm gen}$ and CMRS allocation $m_i(S)$ for agent 1 (left), agent 2 (middle) and agent 3 (right) under the Pareto-Clayton$(0.9)$ model, for $p\in[0.1,0.9]$.}
\label{fig:LRS_CMRS}
\end{figure}

 Figure \ref{fig:LRS_CMRS} compares the simulated VaRs of the initial loss, the equally weighted LRS benchmark,  LRS allocation $A_i^{\mathrm{gen}}$, and the CMRS allocation for all three agents. Observations are summarized below.
\begin{enumerate}[1.]
\item The estimated VaRs under each risk sharing arrangement exceeds that of the initial loss, consistent with the findings in previous sections. The diversification penalty is most pronounced at low probability levels $p$, as shown by the significant gap between the estimated VaRs of $A_i^{\mathrm{eq}}$, $A_i^{\mathrm{gen}}$ and $m_i(S)$, and that of the initial risk $a_iX_i$. This gap gradually decreases as $p$ increases.


\item Relative to the equally weighted benchmark $A_i^{\mathrm{eq}}$, the estimated VaRs under CMRS decrease as the initial exposure $a_i$ increases. The CMRS VaR curve lies above the benchmark for agent 1, close to it for agent 2, and below it for agent 3. This pattern highlights how the two different risk sharing rules are affected by the initial exposures. The equally weighted benchmark $A_i^{\mathrm{eq}}$ scales the same average loss $(X_1+X_2+X_3)/3$ by $a_i$, thus the VaR is directly proportional to each agent's initial exposure $a_i$. On the other hand, under CMRS, each agent's allocation is calculated by averaging its initial loss $a_iX_i$ given the same pool total. When the total pool loss is small, the CMRS allocation for an agent with a small exposure can still include a relatively larger underlying loss $X_i$. For an agent with a larger exposure, the same underlying loss can make the pool total large as well. Hence, a given pool places a tighter range on the underlying loss of an agent with a larger exposure. This may explain why the estimated CMRS VaRs are relatively higher than those of the equally weighted benchmarks $A_i^{\mathrm{eq}}$ for agents with smaller exposures, especially at low probability levels $p$. 

\item A side-by-side comparison of the three agents shows that, for the initial loss $a_iX_i$ and each of the allocations $A_i^{\mathrm{eq}}$, $A_i^{\mathrm{gen}}$, and $m_i(S)$, agent 3 has the highest estimated VaRs, followed by agent 2 and then agent 1. This ranking matches the ordering of their initial exposures with $a_3>a_2>a_1$.
\end{enumerate}

\section{Concluding remarks}
\label{sec:conclusion}
In this paper, we examine stochastic order relations for positively dependent infinite-mean losses via background risk models, under both  linear and conditional mean risk sharing rules. 
While previous studies have mainly focused on the linear risk sharing of independent or negatively dependent risks, we broaden the scope by proving that both risk sharing rules can increase risk in usual stochastic order. Ultimately, under the respective assumptions, these stochastic order relations imply that retaining the original risk or selecting the less diversified portfolio is preferred under every law-invariant monotone risk measure that is well defined for the losses considered. 

For linear risk sharing, Theorem \ref{thm:common-multiplier} connects the inverted concave class for background risks with the concavity of VaR in the exposure vector,  universal VaR superadditivity, and stochastic comparison with exchangeable idiosyncratic risks.  Theorem \ref{thm:radial-archimedean}  establishes a diversification penalty   for models  with inverted concave idiosyncratic risks and $n$-monotone Archimedean copulas.  Theorem \ref{thm:Pareto-mixture} further compares portfolios of Pareto marginals coupled by convex mixtures of several Archimedean copulas. 

For conditional mean risk sharing, Theorem \ref{thm:inverted-concave-common-factor} proves a diversification penalty for multiplicative background risk models with an inverted concave background risk. Moreover, given stable idiosyncratic risks, Theorem \ref{thm: peakedness} and Corollary \ref{cor: CMRS_stochastic} provide peakedness and stochastic order comparisons, respectively, complementing the peakedness order analysis for linear combinations of stable random variables in \citet{ibragimov2005new}. 

Extending the comparisons to more general positive dependence is a natural direction for future research, while a simple example in Remark \ref{positive_dependence_counterexample} has illustrated that, with a general notion of positive dependence, a diversification penalty may fail under linear risk sharing.

\section*{Acknowledgments}
We thank Ruodu Wang for helpful comments on a previous version of the paper.
Ziting Miao is supported by the Australian Government Research Training Program Scholarship and the Henry Buck Scholarship from the University of Melbourne. Yuyu Chen is supported by 2025 FBE ARC and Large External Grant Support Package from the University of Melbourne.

\section*{Declaration of competing interest}
The authors declare that they have no known competing financial interests or personal relationships that could have appeared to influence the work reported in this paper.


\clearpage
\begin{center}
    \Large
    Supplementary Materials for\\
    ``Risk sharing for background risk models with infinite mean''
\end{center}

\vspace{0.1cm}
\setcounter{section}{0}
\renewcommand{\thesection}{\arabic{section}}
\section{Proof of Theorem 3}
\begin{proof}
Define $Z_i=X_i\mid R=r$ for $r>0$. Since $(X_1,\dots,X_n)\laweq(f(Y_1,R),\dots,f(Y_n,R))$, for $z>0$, we have 
\begin{align*}
g_{r}(z)&= \bar{F}_{Z_i}\left(\frac{1}{z}\right)=\p\left(F_{X_i}^{-1}\left(\psi\left(\frac{E_i}{r}\right)\right)>\frac{1}{z}\right)=\p\left(E_i<r\psi^{-1}\left(1-\left(\frac{z}{1+z}\right)^\alpha\right)\right)\\
&=1-\exp\left(-r\psi^{-1}\left(1-\left(\frac{z}{1+z}\right)^\alpha\right)\right).
\end{align*}
By Lemma \ref{thm:conditional-dp}, it suffices to prove $Z_i\in\mathrm{IC}$ for all $r>0$, which is equivalent to verify the concavity of $g_{r}$ for all $r>0$. 
We obtain some auxiliary results. Let $q=\frac{z}{1+z}$, and define $K_\alpha(q)=[(1+2z)(1-q^\alpha)-\alpha]/\alpha q^\alpha$. We will next show that $K_\alpha(q) \geq 1$. For $0<\alpha\le 1$ and $0<q<1$, we have
\[
\begin{aligned}
K_\alpha(q)
&=\frac{(1+2z)(1-q^\alpha)-\alpha}{\alpha q^\alpha}=\frac{1}{\alpha q^\alpha}\frac{1+q}{1-q}(1-q^\alpha)-\frac{1}{q^\alpha}=\frac{1+q^\alpha}{\alpha q^\alpha}\frac{1-q^\alpha}{1+q^\alpha}\frac{1+q}{1-q}-\frac{1}{q^\alpha}\\
&=\frac{1+q^\alpha}{\alpha q^\alpha}\frac{1-e^{-\alpha\log(1/q)}}{1+e^{-\alpha\log(1/q)}}\frac{1+e^{-\log(1/q)}}{1-e^{-\log(1/q)}}-\frac{1}{q^\alpha}=\frac{1+q^\alpha}{\alpha q^\alpha}\frac{\tanh\!\left(\frac{\alpha}{2}\log\frac{1}{q}\right)}{\tanh\!\left(\frac{1}{2}\log\frac{1}{q}\right)}-\frac{1}{q^\alpha}\\
&\ge \frac{1+q^\alpha}{\alpha q^\alpha}\frac{\alpha\tanh\!\left(\frac{1}{2}\log\frac{1}{q}\right)}{\tanh\!\left(\frac{1}{2}\log\frac{1}{q}\right)}-\frac{1}{q^\alpha}=\frac{1+q^\alpha}{q^\alpha}-\frac{1}{q^\alpha}=1.
\end{aligned}
\]
The inequality follows from the concavity of $\tanh$ on $[0,\infty)$, since $\tanh(0)=0$ and $0<\alpha\le 1$.

\begin{enumerate}[(i)]
\item 
\textbf{Independence copula.}
The diversification penalty follows directly from Theorem 1 of \cite{chen2026stochastic}.

\item 
\textbf{Gumbel copula.}
We obtain $g_{r}(z)$ and its second derivative $g_{r}''(z)$ as
\[
g_r(z)=1-\exp\!\left\{-r[-\log(1-q^\alpha)]^\theta\right\},
\]
{\small
\[
g_r''(z)=\frac{r\theta\alpha^2 q^{2\alpha-2}[-\log(1-q^\alpha)]^{\theta-2}}{(1+z)^4(1-q^\alpha)^2}\exp\!\left\{-r[-\log(1-q^\alpha)]^\theta\right\}\left[\theta-1+\log(1-q^\alpha)K_\alpha(q)-r\theta[-\log(1-q^\alpha)]^\theta\right].
\]
}
For $0<\alpha\le 1$ and $z>0$, we have $-\log(1-q^\alpha)=\int_0^{q^\alpha}1/(1-t)\,\mathrm{d}t\ge q^\alpha$.
Since $K_\alpha(q)\ge 1$, it follows that
\[
[-\log(1-q^\alpha)]K_\alpha(q) \geq q^\alpha K_\alpha(q) \geq \frac{q^\alpha+\alpha q^\alpha K_\alpha(q)}{2}=\frac{q^\alpha+(1+2z)(1-q^\alpha)-\alpha}{2} \geq \frac{1-\alpha}{2}.
\]
Hence, for $r>0$, and $\theta\ge 1$, the bound $[-\log(1-q^\alpha)]K_\alpha(q)\ge (1-\alpha)/2$ gives
\[
g_r''(z) \leq \frac{r\theta\alpha^2 q^{2\alpha-2}[-\log(1-q^\alpha)]^{\theta-2}}{(1+z)^4(1-q^\alpha)^2}\exp\!\left\{-r[-\log(1-q^\alpha)]^\theta\right\}\left(\theta-1-\frac{1-\alpha}{2}\right),
\]
where we have dropped the non-positive term $-r\theta[-\log(1-q^\alpha)]^\theta$ from the last bracket. The first two factors are positive for every $z>0$. Thus, to ensure $g_r''(z)\le 0$ for all $z>0$, it suffices to require $\theta-1-(1-\alpha)/2\le 0$. Hence, by fixing $0<\alpha \leq 1$, a sufficient condition is $1\leq\theta\leq (3-\alpha)/2$.

\item
\textbf{Joe copula.}
We obtain $g_{r}(z)$ and its second derivative $g_{r}''(z)$ as
\[
g_r(z)=1-(1-q^{\alpha\theta})^r, \quad g_r''(z)=-\frac{r\alpha\theta q^{\alpha\theta-2}(1-q^{\alpha\theta})^{r-2}}{(1+z)^4}\left[(1+2z)(1-q^{\alpha\theta})+\alpha\theta(rq^{\alpha\theta}-1)\right].
\]
The first factor of $g_r''(z)$ is negative for all $r>0$, $z>0$ and $0<\alpha \leq1$. Suppose that $0<\alpha\theta\le 1$, the concavity of the power function $q\mapsto q^{\alpha\theta}$ and the tangent line at $q=1$ gives $q^{\alpha\theta}\le 1+\alpha\theta(q-1)$. Since $(1+2z)(1-q)=1+q$, the last bracket of $g_r''(z)$ satisfies $(1+2z)(1-q^{\alpha\theta})+\alpha\theta(rq^{\alpha\theta}-1) \geq \alpha\theta\bigl[(1+2z)(1-q)+rq^{\alpha\theta}-1\bigr]=\alpha\theta\bigl(q+rq^{\alpha\theta}\bigr)>0$. Hence, by fixing $0<\alpha \leq 1$, a sufficient condition is $1\leq\theta\leq 1/\alpha$.

\item
\textbf{Clayton copula.}
We obtain $g_{r}(z)$ and its second derivative $g_{r}''(z)$ as
\[
g_r(z)=1-\exp\!\left\{-r\bigl[(1-q^\alpha)^{-\theta}-1\bigr]\right\}, 
\]
\[
g_r''(z)=\frac{r\theta\alpha^2q^{2\alpha-2}}{(1+z)^4(1-q^\alpha)^{\theta+2}}\exp\!\left\{-r\bigl[(1-q^\alpha)^{-\theta}-1\bigr]\right\}\left[\theta-K_\alpha(q)-r\theta(1-q^\alpha)^{-\theta}\right].
\]
For all $r>0$, $\theta>0$ and $0<\alpha \leq1$, the first two factors of $g_r''(z)$ are both positive. Since $0<q<1$, we have $(1-q^\alpha)^{-\theta}\ge 1$. Using $K_\alpha(q)\ge 1$, we obtain $\theta-K_\alpha(q)-r\theta(1-q^\alpha)^{-\theta} \leq \theta-1-r\theta(1-q^\alpha)^{-\theta} \leq (1-r)\theta-1<0$ for all $r>0$ whenever $0<\theta\leq1$. Hence, by fixing $0<\alpha \leq 1$, a sufficient condition is $0<\theta\leq 1$.

\item
\textbf{Ali--Mikhail--Haq copula.}
We obtain $g_{r}(z)$ and its second derivative $g_{r}''(z)$ as
\[
g_r(z)=1-\left(\frac{1-q^\alpha}{1-\theta q^\alpha}\right)^r,
\]
\[
g_r''(z)=\frac{r(1-\theta)\alpha^2 q^{2\alpha-2}}{(1+z)^4(1-q^\alpha)^2(1-\theta q^\alpha)^2}\left(\frac{1-q^\alpha}{1-\theta q^\alpha}\right)^r\left[-(1-\theta q^\alpha)K_\alpha(q)+\theta(1-q^\alpha)-r(1-\theta)\right].
\]
For $0<\alpha\le 1$ and $0\le\theta<1$,
the first two factors of $g_r''(z)$ are positive for every $r>0$ and $z>0$. Since $1-\theta q^\alpha>0$ and $K_\alpha(q)\ge 1$, the last bracket satisfies $-(1-\theta q^\alpha)K_\alpha(q)+\theta(1-q^\alpha)-r(1-\theta) \leq -(1-\theta q^\alpha)+\theta(1-q^\alpha)-r(1-\theta)=-(1+r)(1-\theta)<0$. Hence, by fixing $0<\alpha \leq 1$, a sufficient condition is $0\leq\theta< 1$.

\item
\textbf{Frank copula.}
We obtain $g_{r}(z)$ and its second derivative $g_{r}''(z)$ as
\[
g_r(z)=1-\left(\frac{1-e^{-\theta(1-q^\alpha)}}{1-e^{-\theta}}\right)^r,
\]
\[
g_r''(z)=\frac{r\theta\alpha^2q^{2\alpha-2}e^{-\theta(1-q^\alpha)}}{(1+z)^4(1-e^{-\theta})}\left(\frac{1-e^{-\theta(1-q^\alpha)}}{1-e^{-\theta}}\right)^{r-1}\left[\theta-\frac{K_\alpha(q)+1}{1-q^\alpha}-\frac{(r-1)\theta}{e^{\theta(1-q^\alpha)}-1}\right].
\]
For $0<\alpha\leq 1$, $r>0$, $z>0$, and $\theta>0$, the first two factors of $g_r''(z)$ are positive. The last bracket satisfies
\[
\begin{aligned}
\theta-\frac{K_\alpha(q)+1}{1-q^\alpha}-\frac{(r-1)\theta}{e^{\theta(1-q^\alpha)}-1}
&=\frac{1}{1-q^\alpha}\left[\frac{\theta(1-q^\alpha)}{1-e^{-\theta(1-q^\alpha)}}-K_\alpha(q)-1
\right]-\frac{r\theta}{e^{\theta(1-q^\alpha)}-1}\\
&\leq \frac{1}{1-q^\alpha}\left[\frac{\theta}{1-e^{-\theta}}-2\right]-\frac{r\theta}{e^{\theta(1-q^\alpha)}-1}.
\end{aligned}
\]
The inequality uses $K_\alpha(q)\ge 1$ and the fact that $u\mapsto u/(1-e^{-u})$ is increasing on $(0,\infty)$. Thus, for all all $r>0$, $z>0$ and $0<\theta\leq 1.5$, we have $\theta/(1-e^{-\theta}) \leq 1.5/(1-e^{-1.5})<2$. Hence, by fixing $0<\alpha \leq 1$, a sufficient condition is $0<\theta\leq 1.5$.

\item
\textbf{Survival copula: Clayton.}
When $\theta=1/\alpha$ by fixing $0<\alpha \leq 1$, the model reduces to the Pareto-Clayton$(\alpha)$ model. The diversification penalty follows directly from Example 1.
\end{enumerate}

Since each $g_{r}(z)$ associated with each copula in Table \ref{tab:copula-mixture-examples} is concave under the stated parameter conditions, their convex combination $g_r^C=\sum_{j=1}^{m} p_jg_{r}^j(z)$ for $(p_1,\dots,p_m)\in\Delta_m$ is also concave. The proof is now complete.
\end{proof}

\end{document}